\documentclass[a4paper]{JHEP3}
\usepackage[dvips]{graphicx}
\usepackage{amsfonts}
\usepackage{amssymb}
\usepackage{amsmath}
\usepackage{epsfig}
\usepackage{cite}
\usepackage{multirow}

\newcommand{\be}{\begin{equation}}
\newcommand{\ee}{\end{equation}}
\newcommand{\bea}{\begin{eqnarray}}
\newcommand{\eea}{\end{eqnarray}}
\newcommand{\mbb}{\mathbb}
\newcommand{\ti}{\times}
\newcommand{\half}{\frac{1}{2}}
\newcommand{\mc}{\mathcal}

\newcommand{\beqa}{\begin{eqnarray}}
\newcommand{\eeqa}{\end{eqnarray}}

\newcommand{\ph}{\phantom}

\newcommand{\f}{{\bf 5}}
\newcommand{\fb}{\bar{\bf{5}}}
\newcommand{\te}{{\bf 10}}
\newcommand{\teb}{\bar{\bf{10}}}
\newcommand{\op}{\oplus}

\renewcommand{\o}{\bf{1}}
\newcommand{\nn}{\nonumber}

 \title{Aspects of Flavour and Supersymmetry in F-theory GUTs}
\author{Joseph P. Conlon $^1$, Eran Palti $^2$ \\
 1: Rudolf Peierls Centre for Theoretical Physics, 1 Keble Road, Oxford OX1 3NP, UK \\
 2: Centre de Physique Th´eorique, Ecole Polytechnique, CNRS, 91128 Palaiseau, France. \\
 \\
 E-mail:
  \email{j.conlon1@physics.ox.ac.uk}, \email{palti@cpht.polytechnique.fr}
}

\abstract{We study the constraints of supersymmetry on flavour in recently proposed models
of F-theory GUTs. We relate the topologically twisted theory to the canonical presentation of
eight-dimensional super Yang-Mills and provide a dictionary between the two. We describe the
constraints on Yukawa couplings implied by holomorphy of the superpotential in the
effective 4-dimensional supergravity theory, including the scaling with $\alpha_{GUT}$.
Taking D-terms into account we solve explicitly to second order for wavefunctions and Yukawas due to metric and flux perturbations and
find a rank-one Yukawa matrix with no subleading corrections.}

\preprint{OUTP-09/25P,CPHT-RR107.1009}

\begin{document}

%\tableofcontents

%%%%%%%%%%%%%%%%%%%%%%%%%%%%%%%%%%%%%%%%%%%%%%%%%%%%%%%%%%%%%%%%%%%%%%%%%%%%%
\section{Introduction}
%%%%%%%%%%%%%%%%%%%%%%%%%%%%%%%%%%%%%%%%%%%%%%%%%%%%%%%%%%%%%%%%%%%%%%%%%%%%%

The flavour mass hierarchies of the standard model are a tantalising hint
towards deeper and more fundamental structures in theoretical physics.
As a candidate fundamental theory string theory is a natural arena in which to study this topic.
There has been recent interest in F-theory model building, in part motivated by the ability to
obtain an $\mc{O}(1)$ top Yukawa coupling.
These models were originally proposed in \cite{Donagi:2008ca,Beasley:2008dc} (see \cite{08052943,Hayashi:2008ba,Beasley:2008kw,Donagi:2008kj,Heckman:2008qt,Blumenhagen:2008zz,
Font:2008id,Heckman:2008qa,Blumenhagen:2008aw,09013785,Hayashi:2009ge,Bouchard:2009bu,09024143, 09033009, Donagi:2009ra,Randall:2009dw,09043101,09043932,09052289,09060013,
Marsano:2009gv,Heckman:2009mn,09063297,Conlon:2009qa,Font:2009gq,Blumenhagen:2009yv} for some further developments).
Flavour physics in these models has been studied in \cite{Font:2008id, Heckman:2008qa,  Hayashi:2009ge, Font:2009gq, 09043101,Randall:2009dw, Bouchard:2009bu,
Cecotti:2009zf}.
The models describe locally intersecting 7-branes within a Calabi-Yau four-fold by a Higgsed twisted 8-dimensional gauge theory. The Higgsing induces localised zero modes along curves in the four-dimensional manifold $S$ wrapped by the 7-brane.
In the presence of flux, these zero modes correspond to chiral matter fields in
the four-dimensional effective GUT. The existence of a tree-level top Yukawa coupling
requires the gauge theory to have an underlying exceptional gauge group,
 thereby enforcing the non-perturbative F-theory approach.

Following the initial work an attractive proposal was made in \cite{Heckman:2008qa} where the hierarchical structure of
the Yukawas was proposed to arise from a rank one Yukawa matrix via perturbative corrections in powers of flux.
Furthermore the perturbative parameter was argued to be related to the GUT gauge coupling.
The resulting CKM matrix, following some assumptions regarding order one factors and
the loci where up-type and down-type Yukawas where generated, took a phenomenologically attractive form. This proposal was studied in more detail in subsequent papers \cite{Randall:2009dw, Heckman:2009mn, Font:2009gq}. These papers, as initial studies of flavour, did not deeply explore the effects and constraints of supersymmetry.
The aim of this article is to study the connection between these proposed models of flavour and the constraints arising from supersymmetry.

This paper is organised as follows.
In section \ref{sec2} we review the topologically twisted theory as described in \cite{Beasley:2008dc}.
In section \ref{sec3} we relate this theory to dimensional reduction of
the canonical presentation of 8d super Yang-Mills. We show how the
the equations of motion and Yukawa coupling take an identical form in the two approaches
 and give a dictionary between the two formalisms. We show
 how solutions to the equations of motion are unaltered under overall metric rescalings.
This is important since the volume of $S$ gives the gauge coupling of the effective four-dimensional GUT.
We relate this to the constraints required by holomorphy of the effective 4d supergravity theory
and explain why $\alpha_{GUT}$ must appear universally in the Yukawa couplings.

In section \ref{sec:yukcoup}
we study more specifically the equations of motion.
In particular we study the effect that solving the D-terms has on the form of the localised matter wavefunctions.
For general flux a flat metric on $S$ no longer solves the D-term equations. In order to maintain supersymmetry
the metric is deformed, modifying the equations of motion and subsequently the matter wavefunctions.
We solve for the corrected wavefunctions at leading order in the perturbations.
We find that the metric perturbations dominate over the flux in the wavefunctions and give
the leading corrections to the fluxless flat-metric case.
Finally we use the resulting wavefunctions to calculate the resulting Yukawa couplings. We find that, other than the tree-level top Yukawa, all the Yukawa couplings vanish exactly.

\subsection*{Note added}

While this manuscript was in preparation the paper \cite{Cecotti:2009zf} appeared studying similar issues. There the vanishing of the Yukawa couplings was demonstrated and explained at a deeper level than that which appears in this paper. We refer the reader to
 \cite{Cecotti:2009zf} for a more fundamental explanation for the vanishing of Yukawa couplings.

%%%%%%%%%%%%%%%%%%%%%%%%%%%%%%%%%%%%%%%%%%%%%%%%%%%%%%%%%%%%%%%%%%%%%%%%%%%%%
\section{Local models of intersecting branes in F-theory}
\label{sec2}
%%%%%%%%%%%%%%%%%%%%%%%%%%%%%%%%%%%%%%%%%%%%%%%%%%%%%%%%%%%%%%%%%%%%%%%%%%%%%

In this section we review the topologically twisted theory of \cite{Beasley:2008dc}
that describes local models of intersecting 7-branes in F-theory.
In section \ref{sec3} we shall relate this theory to the direct dimensional reduction of 8-dimensional Super-Yang-Mills (SYM)
and provide a dictionary between the two formulations.

%%%%%%%%%%%%%%%%%%%%%%%%%%%%%%%%%%%%%%%%%%%%%%%%%%%%%%%%%%%%%%%%%%%%%%%%%%%%%
\subsection{Effective theory for intersecting 7-branes}
\label{sec:efffthe}
%%%%%%%%%%%%%%%%%%%%%%%%%%%%%%%%%%%%%%%%%%%%%%%%%%%%%%%%%%%%%%%%%%%%%%%%%%%%%

The relevant theory is an 8-dimensional twisted (off-shell supersymmetric)
gauge theory, with gauge group $G$, describing a 7-brane wrapping a 4-dimensional K\"ahler hypersurface $S$. For a local model $S$ is a shrinkable manifold (more formally it has an ample normal bundle) within a Calabi-Yau (CY) 4-fold $X$. The 8-dimensional fields are given in terms of adjoint valued, $S$-valued, 4-dimensional $N=1$ multiplets
\bea
{\bf A}_{\bar{m}} &=& \left( A_{\bar{m}}, \psi_{\bar{m}}, G_{\bar{m}}\right) \;, \\
{\bf \Phi}_{mn} &=& \left( \varphi_{mn}, \chi_{mn}, H_{mn} \right) \;, \\
{\bf V} &=& \left( \eta, A_{\mu}, D \right) \;.
\eea
The indices on the fields denote their form-values on $S$. So for example $A_{\bar{m}} \in \bar{\Omega}^1_{S}\otimes\mathrm{ad}(P)$ where $\Omega^p_S$ denotes holomorphic $p$-form on $S$ and $P$ is the principle bundle (in the adjoint representation) associated to the gauge group $G$.
Here ${\bf A}$ and ${\bf \Phi}$ are chiral multiplets with respective F-terms $G$ and $H$. ${\bf V}$ is a vector multiplet with D-term $D$. $A_{\bar{m}}$ and $\varphi_{mn}$ are complex scalars while $\psi_{\bar{m}}$, $\chi_{mn}$, and $\eta$ are fermions.

The action for the effective theory was given in \cite{Beasley:2008dc}.
Setting 4-dimensional variations of the fields to zero, the equations of motion that follow are
\bea
& &H - F^{(2,0)} = 0 \;, \label{eom1} \\
& &i\left[ \varphi, \bar{\varphi} \right] + 2\omega \wedge F^{(1,1)} + \star_{S} D = 0 \;, \label{eom2} \\
& &2i\omega \wedge \bar{G} - \bar{\partial}_{A} \varphi = 0 \;, \label{eom3} \\
& &- \partial \bar{H} + 2 \omega \wedge \bar{\partial} D + \bar{G} \wedge \bar{\varphi} - \bar{\chi} \wedge \bar{\psi}- i 2\sqrt{2} \omega \wedge \eta \wedge \psi = 0 \;,  \label{eom4} \\
& &\omega \wedge \partial_{A} \psi + \frac{i}{2} \left[ \bar{\varphi}, \chi \right] = 0 \;,  \label{eom5} \\
& &\bar{\partial}_A \chi - 2i\sqrt{2} \omega \wedge \partial_A \eta - \left[ \varphi, \psi \right] = 0 \;,  \label{eom6} \\
& &\bar{\partial}_{A} \psi - \sqrt{2} \left[ \bar{\varphi}, \eta\right] = 0 \;, \label{eom7} \\
& &-\sqrt{2}\left[\bar{\eta},\bar{\chi}\right] - \bar{\partial}_{A} G - \frac12 \left[\psi,\psi\right]= 0 \;. \label{eom8}
\eea
These are the equations that we work with in this paper and they apply to a general K\"ahler manifold $S$ of large enough volume to neglect $\alpha'$ corrections. Equation (\ref{eom1}) imposes that the flux must be of type $(1,1)$. Equation (\ref{eom2}) is the D-term equation that we discuss in detail in section \ref{sec:fluxd}. In this paper we are concerned with vacua where the vacuum expectation value (vev) of $\varphi$, denoted $\left<\varphi\right>$, and that of $A_{\bar{m}}$, which is responsible for the flux, take values in the Cartan of $G$. Therefore equation (\ref{eom3}) just imposes that $\left<\varphi\right>$ is holomorphic
\be
\bar{\partial}_{A} \left<\varphi\right> = \bar{\partial} \left<\varphi\right> + \left[ A , \left<\varphi\right> \right] = \bar{\partial} \left<\varphi\right>  = 0 \;.
\ee
We are interested in solving the equations on a vacuum with
\be
\left< \chi \right> = \left< \psi \right> = \left< \eta \right> = 0 \;, \label{vanishvev}
\ee
which means that, using equations (\ref{eom1}-\ref{eom3}) equations (\ref{eom4}) and (\ref{eom8}) are satisfied.
Equation (\ref{eom7}) follows from (\ref{eom6}) once (\ref{vanishvev}) is imposed.

We take the manifold $S$ to be K\"ahler and spanned by two complex coordinates $z_1$ and $z_2$. We will also restrict to the metric ansatz
\be
ds^2 = h_1\left(z_1, \bar{z}_{\bar{1}}\right) \; dz_1 \otimes d \bar{z}_{\bar{1}} + h_2\left(z_2, \bar{z}_{\bar{2}}\right) \;dz_2 \otimes d \bar{z}_{\bar{2}} \;, \label{metricans}
\ee
where $h_1$ and $h_2$ are general functions of $z_1$ and $z_2$ respectively (this is the form of the metric we will use
when solving for explicit wavefunctions in section \ref{sec:yukcoup}).
The corresponding K\"ahler form is given by
\be
\omega = \frac{i}{2} h_1 dz_1 \wedge d\bar{z}_{\bar{1}} + \frac{i}{2} h_2 dz_2 \wedge d\bar{z}_{\bar{2}} \;.
\ee
Putting this into (\ref{eom1}-\ref{eom8}) the relevant equations read
\bea
& & h_1 \partial_{A2} \psi_{\bar{2}} + h_2 \partial_{A1} \psi_{\bar{1}} - \left[ \bar{\varphi}, \chi \right] = 0\;, \label{preqofmot1} \\
& & \bar{\partial}_{A\bar{1}} \chi - \left[ \varphi, \psi_{\bar{1}} \right]  = 0 \;, \label{preqofmot2} \\
& & \bar{\partial}_{A\bar{2}} \chi - \left[ \varphi, \psi_{\bar{2}} \right]  = 0 \;, \label{preqofmot3}
\eea
where by abuse of notation we have relabeled $\varphi_{12} \rightarrow \varphi$ and $\chi_{12} \rightarrow \chi$. Apart from these equations
we also have the D-term equation (\ref{eom2}).

%%%%%%%%%%%%%%%%%%%%%%%%%%%%%%%%%%%%%%%%%%%%%%%%%%%%%%%%%%%%%%%%%%%%%%%%%%%%%
\subsection{Zero modes and Yukawa couplings}
\label{sec:zeroyuk}
%%%%%%%%%%%%%%%%%%%%%%%%%%%%%%%%%%%%%%%%%%%%%%%%%%%%%%%%%%%%%%%%%%%%%%%%%%%%%

Consider turning off the flux so that there is no gauge field background. We then turn on a vev for $\varphi$ given by
\be
\left< \varphi \right> = m_1^2 z_1 Q_1 + m_2^2 z_2 Q_2 \;. \label{vevphi}
\ee
Here $Q_1$ and $Q_2$ are elements in $G$. In this paper we perform the simplification $m_1=m_2=m$ and define
\be
v \equiv \frac{1}{m^2}\;.
\ee
Since there is no flux the D-term equation is solved identically and we are free to consider a flat metric background $h_1=h_2=1$.
The equations of motion to be solved then read
\bea
& &\partial_2 \psi_{\bar{2}} + \partial_1 \psi_{\bar{1}} -  \frac{1}{v}\left( \bar{z}_1 q_1 + \bar{z}_2 q_2\right) \chi = 0\;, \label{0oeqofmot1} \\
& & \bar{\partial}_{\bar{1}} \chi - \frac{1}{v} \left( z_1 q_1 + z_2 q_2\right) \psi_{\bar{1}} = 0 \;, \label{0oeqofmot2} \\
& & \bar{\partial}_{\bar{2}} \chi - \frac{1}{v} \left( z_1 q_1 + z_2 q_2\right) \psi_{\bar{2}} = 0 \;. \label{0oeqofmot3}
\eea
Here $q_1$ and $q_2$ are the charges of the fields under $Q_1$ and $Q_2$. For simplicity consider the case $q_1=1$ and $q_2=0$. Then the equations have a localised solution given by
\be
\chi = f(z_2) e^{-\frac{|z_1|^2}{v}} \;.
\ee
Here and in section \ref{sec:yukcoup}, when presenting a solution we only present the expression for $\chi$ since the expressions for $\psi_{\bar{1}}$ and $\psi_{\bar{2}}$ can be determined directly from it using (\ref{0oeqofmot2}) and (\ref{0oeqofmot3}).
This is a localised zero mode which can be thought of as strings stretching between two 7-branes that are intersecting along $z_1=0$. If we decompose the 8-dimensional fields into external space-time 4-dimensional components and internal 4-dimensional components along $S$ we see that the equations imply that the external 4-dimensional components of $\chi$, $\psi_{\bar{1}}$ and $\psi_{\bar{2}}$ are all equal and they count as a single 4-dimensional zero mode.

If flux is turned on there are two changes: the form of the internal wavefunctions changes and the number of zero modes can change. The change in the form of the wavefunctions is a local effect that we calculate explicitly in section \ref{sec:yukcoup}. The replication of zero modes is of course the family index and depends on the global properties of the flux. In a local model we decouple the connection between the flux and the zero mode replication.

The Yukawa couplings are associated to wavefunction overlaps on a point of intersection of three curves where the gauge symmetry is enhanced. For example the bottom type Yukawas are associated to a point of $SO(12)$ gauge symmetry. The matter arises from the decomposition of the adjoint
\bea
SO(12) &\supset& SU(5) \times U(1)_a \times U(1)_b \;, \\
\bf{66} &\rightarrow& \bf{24}^{(0,0)} \op \o^{(0,0)} \op \o^{(0,0)} \op \left(\f \op \fb\right)^{(-1,0)} \op \left(\f \op \fb\right)^{(1,1)} \op
\left(\te \op \teb\right)^{(0,1)} \;, \nn
\eea
where the superscripts denote the charges of the $\f$ and $\te$. Then we see that the $\fb_H \fb_M \te_M$ Yukawa, corresponds to curves of charges
$(-1,-1)$, $(1,0)$ and $(0,1)$ respectively. We label the matter from each of the three curves with the index $\lambda=1,2,3$ corresponding to $\fb_H$, $\fb_M$ and $\te_M$ respectively. The fields are therefore the sum over the localised solutions
\be
\chi = \sum_\lambda \chi^{\lambda} \;, \;\; \psi = \sum_\lambda \psi^{\lambda} \;.
\ee
Since the equations of motion are linear they exactly decouple and can be solved separately for each component.

There is also a family index $a=1,2,3$ which counts copies of the zero modes associated to global properties of the flux. A four-dimensional $\phi$ field is then labeled by
\be
\phi^{\left(\lambda,a\right)} : \; \left\{ \chi^{\left(\lambda,a\right)}, \psi_{\bar{1}}^{\left(\lambda,a\right)}, \psi_{\bar{2}}^{\left(\lambda,a\right)} \right\} \;. \label{4dfieldind}
\ee
Since there is only one Higgs generation for $\lambda=1$ we only have $a=1$.

The relevant Yukawa coupling are the sum over all the terms arising from
\be
\int_{S} A \wedge A \wedge \Phi = -\int_S \hat{\psi}_{\bar{1}}^i \hat{\psi}_{\bar{2}}^j \hat{\chi}_{12}^k 2\mathrm{Tr} \left( \left[ t_i, t_j\right] t_k\right) dz_1 \wedge d\bar{z}_{\bar{1}} \wedge dz_2 \wedge d\bar{z}_{\bar{2}} \;. \label{rawyukawa}
\ee
Here $t_i$ are generators of the point gauge group $SO(12)$. Under the decomposition these correspond to curve indices, $\lambda$ in (\ref{4dfieldind}), with
\be
\lambda=1 :\; \f_{H} \;,\;\;\; \lambda=2 :\; \f_{M} \;,\;\;\; \lambda=3 :\; \te_{M} \;.
\ee
We therefore obtain the Yukawas
\bea
Y^{ab} &=& a \int_S \Bigg[ \hat{\psi}_{\bar{1}}^{1}\hat{\psi}_{\bar{2}}^{3,b}\hat{\chi}^{2,a}  +  \hat{\psi}_{\bar{1}}^{2,a}\hat{\psi}_{\bar{2}}^{1}\hat{\chi}^{3,b}  -  \hat{\psi}_{\bar{1}}^{2,a}\hat{\psi}_{\bar{2}}^{3,b}\hat{\chi}^{1} -  \hat{\psi}_{\bar{1}}^{1}\hat{\psi}_{\bar{2}}^{2,a}\hat{\chi}^{3,b}  - \nonumber \\
 & & \hat{\psi}_{\bar{1}}^{3,b}\hat{\psi}_{\bar{2}}^{1}\hat{\chi}^{2,a}  + \hat{\psi}_{\bar{1}}^{3,b}\hat{\psi}_{\bar{2}}^{2,a}\hat{\chi}^{1}  \Bigg]  \epsilon  \;. \label{yukcompute}
\eea
Here $\epsilon$ is the canonical volume form as in (\ref{rawyukawa}). The constant $a$ is a normalisation factor which is used to normalise the top
Yukawa to 1.

\section{Relationship to Super Yang-Mills}
\label{sec3}

Above we have reviewed, following
\cite{Beasley:2008dc}, the topological equations for the dynamics of the 8-dimensional field theory.
In this section we show how these equations arise from direct analysis of
the canonical 8d SYM Lagrangian and provide a dictionary that translates between the fields of the canonical presentation
of 8d SYM and the fields of the topologically twisted theory used in \cite{Beasley:2008dc}. While ultimately the equations are identical,
this provides an alternative perspective on the topological theory
of \cite{Beasley:2008dc} and may make certain properties of the theory more intuitive.

8d super Yang-Mills is most easily obtained by dimensional
reduction of the 10d theory. The 10d action is
\be
\int \sqrt{g} d^{10} x  \left( - \frac{1}{4} \hbox{Tr} \left( F_{MN} F^{MN} \right) - \frac{1}{2}
\hbox{Tr}\left(\bar{\lambda} \Gamma^M \mc{D}_M \lambda \right) \right),
\ee
where $\lambda$ is a Majorana-Weyl spinor and $\mc{D}_M X= \partial_m X -i [A_m, X]$.
Dimensional reduction gives the 8-dimensional action,
\be
\int d^8 x \sqrt{g} \left( - \frac{1}{4} \hbox{Tr}(F_{MN} F^{MN}) - \frac{1}{4} \hbox{Tr}(\mc{D}_M \phi \mc{D}^M \phi^{*}) + \hbox{quartics}
- \half \hbox{Tr}( \bar{\lambda} \Gamma^m D_m \lambda) + \frac{i}{2} \bar{\lambda} \Gamma^r [\phi_r, \lambda] \right)
\ee
We drop quartic scalar interactions as they are not relevant for our purposes.

Let us now describe $\lambda$. It is convenient
to view $\lambda$ as a 10-dimensional Majorana-Weyl spinor with 8 complex degrees of
freedom (before imposing equations of motion).
We follow \cite{08070789} to write the flat space gamma matrices as $(\mu = 0,1,2,3, m=4,5,6,7, r=8,9)$
$$
\Gamma^\mu = \gamma^\mu \otimes \mbb{I} \otimes \mbb{I}, \qquad \Gamma^m = \gamma^5 \otimes \tilde{\gamma}^{m} \otimes \mbb{I}, \qquad \Gamma^r = \gamma^5 \otimes \tilde{\gamma}^5 \otimes \tau^r,
$$
with
$$
\gamma^0 = \left( \begin{array}{cc} 0 & - \mbb{I}  \\ \mbb{I} & 0 \end{array} \right),
\quad
\gamma^1 = \left( \begin{array}{cc} 0 & \sigma_x \\ \sigma_x & 0 \end{array} \right),
\quad
\gamma^2 = \left( \begin{array}{cc} 0 & \sigma_y \\ \sigma_y & 0 \end{array} \right),
\quad
\gamma^3 = \left( \begin{array}{cc} 0 & \sigma_z \\ \sigma_z & 0 \end{array} \right),
\quad
\gamma^5 = \left( \begin{array}{cc} 1 & 0 \\ 0 & -1 \end{array} \right),
$$
$$
\tilde{\gamma}^1 = \left( \begin{array}{cc} 0 & -i \mbb{I} \\ i\mbb{I} & 0 \end{array} \right),
\quad
\tilde{\gamma}^2 = \left( \begin{array}{cc} 0 & \sigma_z \\ \sigma_z & 0 \end{array} \right),
\quad
\tilde{\gamma}^3 = \left( \begin{array}{cc} 0 & \sigma_x \\ \sigma_x & 0 \end{array} \right),
\quad
\tilde{\gamma}^4 = \left( \begin{array}{cc} 0 & \sigma_y \\ \sigma_y & 0 \end{array} \right),
\quad
\tilde{\gamma}^5 = \left( \begin{array}{cc} \mbb{I} & 0 \\ 0 & \mbb{I} \end{array} \right),
\quad
$$
and
$$
\tau^8 = \sigma_x = \left( \begin{array}{cc} 0 & 1 \\ 1 & 0 \end{array} \right), \quad
\tau^9 = \sigma_y = \left( \begin{array}{cc} 0 & -i \\ i & 0 \end{array} \right).
$$
Using $u$ to denote the complexified 8,9 directions,
we also note
$$
\tau^u \equiv \sigma_x + i \sigma_y = 2  \left( \begin{array}{cc} 0 & 1 \\ 0 & 0 \end{array} \right), \quad
\tau^{\bar{u}} \equiv \sigma_x - i \sigma_y = 2  \left( \begin{array}{cc} 0 & 0 \\ 1 & 0 \end{array} \right), \quad
$$
Spinors can be decomposed by their chiralities under each of the
three (4d + 4d + 2d) components of the gamma matrices.
The Majorana condition can be written as $\lambda^* = B \lambda$, where $B = \Gamma^2 \Gamma^4 \Gamma^7 \Gamma^9$.

As in \cite{08070789} it can be shown that a general Majorana-Weyl spinor can be
written as
\be
\label{EqSpin}
\lambda = (\lambda_1 + \lambda_4) \oplus (\lambda_2 + \lambda_3)
\ee
where
$(\lambda_1, \lambda_2, \lambda_3, \lambda_4) = \{ \lambda^{ab\alpha}, \lambda^a_{\ph{a}b\alpha}, \lambda_{ab}^{\ph{a} \ph{b} \alpha}, \lambda_{a\phantom{b}\alpha}^{\phantom{a}b})$. Super/sub scripts indicate positive or negative chirality in that direction. The notation $(\lambda_1
+ \lambda_4)$ indicates that the components of $\lambda_1$ and $\lambda_4$ are not independent: they are related by the Majorana condition.
More precisely, we have
\bea
\label{Majrelats}
\lambda_1 + \lambda_4 & = &
\left( \begin{array}{c} \xi_1 \\ 0 \end{array} \right) \otimes \left( \begin{array}{c} \psi_1 \\ 0 \end{array} \right) \otimes
\left( \begin{array}{c} \theta_1 \\ 0 \end{array} \right)  +
\left( \begin{array}{c} 0 \\ - \sigma_y \xi_1^{*} \end{array} \right) \otimes \left( \begin{array}{c} -i \sigma_y \psi_1^{*} \\ 0 \end{array} \right) \otimes
\left( \begin{array}{c} 0 \\ -i \theta_1^{*} \end{array} \right), \nonumber \\
\lambda_2 + \lambda_3 & = &
\left( \begin{array}{c} \xi_2 \\ 0 \end{array} \right) \otimes \left( \begin{array}{c} 0 \\ \psi_2 \end{array} \right) \otimes
\left( \begin{array}{c} 0 \\ \theta_2 \end{array} \right)  +
\left( \begin{array}{c} 0 \\ - \sigma_y \xi_2^{*} \end{array} \right) \otimes \left( \begin{array}{c} 0 \\ -i \sigma_y \psi_2^{*} \end{array} \right) \otimes
\left( \begin{array}{c} i \theta_2^{*} \\ 0 \end{array} \right).
\label{lambdas}
\eea
$\lambda_4$ and $\lambda_3$ correspond to the CPT conjugates of $\lambda_1$ and $\lambda_2$ respectively, and thus do not represent
physically distinct degrees of freedom.

For the extra-dimensional spinor components we will also often write
\be
\left( \begin{array}{c} \psi_1 \\ \psi_2 \end{array} \right) \to \left( \begin{array}{c}
\psi_{1,1} \\ \psi_{1,2} \\ \psi_{2,1} \\ \psi_{2,2} \end{array} \right).
\ee

\subsection{Equations of Motion}

We need
to account for the non-trivial metric on the surface wrapped by the branes.
For this we suppose we start with a metric of the factorised form
\be
ds^2 = 4h_1(z, \bar{z}) dz d \bar{z} + 4 h_2(w, \bar{w}) dw d \bar{w},
\ee
so $g_{z \bar{z}} = 2 h_1(z, \bar{z}), g_{w \bar{w}} = 2 h_2(w, \bar{w}), g^{z \bar{z}} =
\frac{1}{2 h_1(z, \bar{z})}, g^{w \bar{w}} = \frac{1}{2 h_2(w, \bar{w})}$.
The spin connection is given by
\bea
\omega_{\mu}^{ab} & = & \half e^{a \nu} \left( \partial_{\mu} e_{\nu}^b -
\partial_{\nu} e_{\mu}^b \right) - \half e^{b \nu} \left( \partial_{\mu} e^a_{\nu}
- \partial_{\nu} e^a_{\mu} \right) - \half e^{\psi a} e^{\sigma b} \left( \partial_{\psi}
e_{\sigma c} - \partial_{\sigma} e_{\psi c} \right) e^c_{\mu}. \, \,
\eea
Here $a,b$ are the vielbein indices with a flat metric while $\mu, \nu$ are the spacetime indices
that run over $z,\bar{z},w, \bar{w}$.
As a vielbein we take
$$
e^1_z = h_1^{\half}, \qquad e^1_{\bar{z}} = h_1^{\half}, \qquad e^1_w = 0, \qquad e^1_{\bar{w}} = 0.
$$
$$
e^2_z = ih_1^{\half}, \qquad e^2_{\bar{z}} = -i h_1^{\half}, \qquad e^2_w = 0, \qquad e^2_{\bar{w}} = 0.
$$
$$
e^3_z = 0, \qquad e^3_{\bar{z}} = 0, \qquad e^3_w = h_2^{\half}, \qquad e^3_{\bar{w}} = h_2^{\half}.
$$
$$
e^4_z = 0, \qquad e^4_{\bar{z}} = 0, \qquad e^4_w = i h_2^{\half}, \qquad e^4_{\bar{w}} = -i h_2^{\half}.
$$
From these we can compute the non-vanishing elements of the spin connection to be
$$
\omega_z^{12} = - \frac{i}{2h_1} \partial_z h_1, \qquad \omega_{\bar{z}}^{12} = \frac{i}{2h_1} \partial_{\bar{z}} h_1,
\qquad \omega_w^{34} = -\frac{i}{2h_2} \partial_w h_2, \qquad \omega_{\bar{w}}^{34} = \frac{i}{2h_2} \partial_{\bar{w}} h_2.
$$
We also need to determine the gamma matrices. We have
\be
\tilde{\gamma}^z = e^{z1} \tilde{\gamma}_1 + e^{z2} \tilde{\gamma}_2 = \frac{h_1^{-1/2}}{2}
\left( \tilde{\gamma}_1 - i \tilde{\gamma}_2 \right), \qquad
\tilde{\gamma}^{\bar{z}} = e^{\bar{z}1} \tilde{\gamma}_1 + e^{\bar{z}2} \tilde{\gamma}_2 =
\frac{h_1^{-1/2}}{2} \left( \tilde{\gamma}_1 + i \tilde{\gamma}_2 \right),
\ee
\be
\tilde{\gamma}^w = \frac{h_2^{-\half}}{2} \left( \tilde{\gamma}_3 - i \tilde{\gamma}_4 \right), \qquad
\tilde{\gamma}^{\bar{w}} = \frac{h_2^{-\half}}{2} \left( \tilde{\gamma}_3 + i \tilde{\gamma}_4 \right).
\ee
Therefore we can write
\bea
\tilde{\gamma}^z   =  \frac{h_1^{-\half}}{2} \left( \begin{array}{cccc} 0 & 0 & -2i & 0 \\ 0 & 0 & 0 & 0 \\
0 & 0 & 0 & 0 \\ 0 & 2i & 0 & 0 \end{array} \right), \quad & &
\tilde{\gamma}^{\bar{z}} = \frac{h_1^{-\half}}{2} \left( \begin{array}{cccc} 0 & 0 & 0 & 0 \\ 0 & 0 & 0 & -2i \\
2i & 0 & 0 & 0 \\ 0 & 0 & 0 & 0 \end{array} \right), \nonumber \\
\tilde{\gamma}^{w}  =  \frac{h_2^{-\half}}{2} \left( \begin{array}{cccc} 0 & 0 & 0 & 0 \\ 0 & 0 & 2 & 0 \\
0 & 0 & 0 & 0 \\ 2 & 0 & 0 & 0 \end{array} \right), \quad & &
\tilde{\gamma}^{\bar{w}} = \frac{h_2^{-\half}}{2} =  \left( \begin{array}{cccc} 0 & 0 & 0 & 2 \\ 0 & 0 & 0 & 0 \\
0 & 2 & 0 & 0 \\ 0 & 0 & 0 & 0 \end{array} \right).
\eea

For the spin connection we need
$\frac{1}{8} \gamma^M \omega_{M, \alpha \beta} \left[ \gamma^{\alpha}, \gamma^{\beta} \right]$. As
$
\left[ \gamma^1, \gamma^2 \right] = -2i \left( \begin{array}{cc} \sigma_z & 0 \\ 0 & -\sigma_z \end{array} \right)$,
$\left[ \gamma^3, \gamma^4 \right] = 2i \left( \begin{array}{cc} \sigma_z & 0 \\ 0 & \sigma_z \end{array} \right)$,
we have
$$
\frac{1}{8} \omega_{z \alpha \beta} \left[ \gamma^{\alpha}, \gamma^{\beta} \right] = -\frac{\partial_z h_1}{4h_1}
\left( \begin{array}{cc} \sigma_z & 0 \\ 0 & -\sigma_z \end{array} \right), \qquad
\frac{1}{8} \omega_{\bar{z} \alpha \beta} \left[ \gamma^{\alpha}, \gamma^{\beta} \right] = \frac{\partial_{\bar{z}} h_1}{4h_1}
\left( \begin{array}{cc} \sigma_z & 0 \\ 0 & -\sigma_z \end{array} \right).
$$
$$
\frac{1}{8} \omega_{w \alpha \beta} \left[ \gamma^{\alpha}, \gamma^{\beta} \right] = \frac{\partial_w h_2}{4h_2}
\left( \begin{array}{cc} \sigma_z & 0 \\ 0 & \sigma_z \end{array} \right), \qquad
\frac{1}{8} \omega_{\bar{w} \alpha \beta} \left[ \gamma^{\alpha}, \gamma^{\beta} \right] = \frac{\partial_{\bar{w}} h_2}{4h_2}
\left( \begin{array}{cc} \sigma_z & 0 \\ 0 & \sigma_z \end{array} \right).
$$
Putting these together we obtain
\be
\gamma^M \partial_M + \frac{1}{4} \gamma^M \omega_{M \alpha \beta} \gamma^{\alpha \beta} =
\left( \begin{array}{cccc} 0 & 0 & -\frac{i}{h_1^{\half}} \left( \partial_z + \frac{\partial_z h_1}{4h_1} \right)
& \frac{1}{h_2^{\half}} \left( \partial_{\bar{w}} + \frac{\partial_{\bar{w}} h_2}{4 h_2} \right) \\
0 & 0 & \frac{1}{h_2^{\half}} \left( \partial_w + \frac{\partial_w h_2}{4h_2} \right) & \frac{-i}{h_1^{\half}} \left(
\partial_{\bar{z}} + \frac{\partial_{\bar{z}} h_1}{4 h_1} \right) \\
\frac{i}{h_1^{\half}} \left( \partial_{\bar{z}} + \frac{\partial_{\bar{z}} h_1}{4h_1} \right) & \frac{1}{h_2^{\half}} \left(
\partial_{\bar{w}} + \frac{\partial_{\bar{w}} h_2}{4h_2} \right) & 0 & 0 \\
\frac{1}{h_2^{\half}} \left( \partial_w + \frac{\partial_w h_2}{4h_2} \right) & \frac{i}{h_1^{\half}} \left( \partial_z
+ \frac{\partial_z h_1}{4h_1} \right) & 0 & 0
\end{array} \right),
\ee
where $\gamma^{\alpha \beta} \equiv \half [\gamma^{\alpha}, \gamma^{\beta}]$.
We need to twist this operator to account for the nontrivial metric. The twisting corresponds
to an additional gauge field in the canonical bundle. To determine this, let us first recall that a metric
\be
ds^2 = 4 h_1(z, \bar{z}) dz d \bar{z} + 4 h_2(w, \bar{w}) dw d \bar{w},
\ee
has
$$
\Gamma^z_{zz} = \frac{\partial_z h_1}{h_1}, \qquad \Gamma^{\bar{z}}_{\bar{z} \bar{z}} =
\frac{\partial_{\bar{z}} h_1}{h_1}, \qquad \Gamma^w_{ww} = \frac{\partial_w h_2}{h_2}, \qquad
\Gamma^{\bar{w}}_{\bar{w} \bar{w}} = \frac{\partial_{\bar{w}} h_2}{h_2}.
$$
The twisting corresponds to a gauge field in the canonical bundle. A gauge field
with $M$ units of flux in the $z\bar{z}$ direction and $N$ units of flux in the $w\bar{w}$ direction has
$$
A_z = - \frac{i M \partial_z h_1}{4h_1}, \qquad A_{\bar{z}} = \frac{i M \partial_{\bar{z}} h_1}{4h_1},
\qquad A_w = - \frac{i N \partial_w h_2}{4h_2}, \qquad A_{\bar{w}} = \frac{i N \partial_{\bar{w}} h_2}{4h_2}.
$$
The factor of $i/4$ can be determined by reference to the case of $\mbb{P}^1 \ti \mbb{P}^1$, for which $z$ and $w$
parameterise the two separate $\mbb{P}^1$s. This gives
\be
\gamma^M \partial_M + \frac{1}{4} \gamma^M \omega_{M \alpha \beta} \gamma^{\alpha \beta} - i \gamma^M A_M =
\ee
\be
\label{fulldirac}
\left( \begin{array}{cccc} 0 & 0 & \frac{-i}{h_1^{\half}} \left( \partial_z + \frac{\partial_z h_1}{4h_1}(1-M) \right) &
\frac{1}{h_2^{\half}}\left( \partial_{\bar{w}} + \frac{\partial_{\bar{w}} h_2}{4 h_2} \left( 1 + N \right) \right) \\
0 & 0 & \frac{1}{h_2^{\half}} \left( \partial_w + \frac{\partial_w h_2}{4h_2} (1-N) \right) & \frac{-i}{h_1^{\half}} \left(
\partial_{\bar{z}} + \frac{\partial_{\bar{z}} h_1}{4 h_1}(1+M) \right) \\
\frac{i}{h_1^{\half}} \left( \partial_{\bar{z}} + \frac{\partial_{\bar{z}} h_1}{4h_1}(1+M) \right) & \frac{1}{h_2^{\half}} \left(
\partial_{\bar{w}} + \frac{\partial_{\bar{w}} h_2}{4h_2}(1+N) \right) & 0 & 0 \\
\frac{1}{h_2^{\half}} \left( \partial_w + \frac{\partial_w h_2}{4h_2}(1-N) \right) & \frac{i}{h_1^{\half}} \left( \partial_z
+ \frac{\partial_z h_1}{4h_1}(1-M) \right) & 0 & 0
\end{array} \right) \nonumber
\ee
As discussed in \cite{Beasley:2008dc, 08070789}, the left- and right-handed spinors $\psi_1$ and $\psi_2$ have opposite R-charges and so should be
twisted in opposite directions. To accomplish this we modify the Dirac operator (\ref{fulldirac}) by
$$
(M,N) \to (M,N) + (1,1)  \qquad \hbox{ for } (\psi_{1,1}, \psi_{1,2}),
$$
$$
(M,N) \to (M,N) - (1,1)  \qquad \hbox{ for } (\psi_{2,1}, \psi_{2,2}).
$$
The twisted Dirac operator is then
\be
\gamma^M \partial_M + \frac{1}{4} \gamma^M \omega_{M \alpha \beta} \gamma^{\alpha \beta} - i \gamma^M A_{twisted,M} =
\ee
\be
\label{fulldirac2}
\left( \begin{array}{cccc} 0 & 0 & \frac{-i}{h_1^{\half}} \left( \partial_z + \frac{\partial_z h_1}{4h_1}(2-M) \right) &
\frac{1}{h_2^{\half}}\left( \partial_{\bar{w}} + N\frac{\partial_{\bar{w}} h_2}{4 h_2} \right) \\
0 & 0 & \frac{1}{h_2^{\half}} \left( \partial_w + \frac{\partial_w h_2}{4h_2} (2-N) \right) & \frac{-i}{h_1^{\half}} \left(
\partial_{\bar{z}} + M\frac{\partial_{\bar{z}} h_1}{4 h_1} \right) \\
\frac{i}{h_1^{\half}} \left( \partial_{\bar{z}} + \frac{\partial_{\bar{z}} h_1}{4h_1}(2+M) \right) & \frac{1}{h_2^{\half}} \left(
\partial_{\bar{w}} + \frac{\partial_{\bar{w}} h_2}{4h_2}(2+N) \right) & 0 & 0 \\
\frac{1}{h_2^{\half}} \left( \partial_w - N\frac{\partial_w h_2}{4h_2} \right) & \frac{i}{h_1^{\half}} \left( \partial_z
- M\frac{ \partial_z h_1}{4h_1} \right) & 0 & 0
\end{array} \right). \nonumber
\ee
For compactness we write
\be
\label{abcd}
\gamma^M \partial_M + \frac{1}{4} \gamma^M \omega_{M \alpha \beta} \gamma^{\alpha \beta} - i \gamma^M A_{twisted,M}
\equiv \gamma^M \mc{D}_M
\equiv \left( \begin{array}{cc} 0 & \mc{D}_{-} \\ \mc{D}_{+} & 0 \end{array} \right).
\ee
Note that for $M=N=0$, $\psi_{2,2} = 1$ is a solution to $\gamma^M \mc{D}_M \psi = 0$. This corresponds to the
constant gaugino zero mode which should exist for arbitrarily curved spaces. The general solutions
are
\be
\left( \begin{array}{c} \psi_{1,1} \\ \psi_{1,2} \\ \psi_{2,1} \\ \psi_{2,2} \end{array} \right)
= \left( \begin{array}{c} h_1^{-\frac{M+2}{4}} h_2^{\frac{N}{4}} \psi(z, \bar{w}) \\ h_1^{\frac{M}{4}} h_2^{-\frac{N+2}{4}} \psi(\bar{z}, w)
\\ h_1^{\frac{M-2}{4}}h_2^{\frac{N-2}{4}} \psi(\bar{z}, \bar{w}) \\ h_1^{-\frac{M}{4}} h_2^{-\frac{N}{4}} \psi(z, w) \end{array}
\right).
\ee

To study, at least locally, intersecting branes
we also need the profile for the Higgs field, namely the complex
scalar $\phi$ that is present in the 8d SYM theory. In an untwisted background this field would satisfy
$$
\partial_{\bar{z}} \phi_u =0, \quad \partial_{z} \phi_{\bar{u}} = 0, \quad \partial_{\bar{w}} \phi_u = 0,
\quad \partial_{w} \phi_{\bar{u}} = 0.
$$
Zero mode solutions are therefore given by
$$
\mc{D}_{z} \phi_{\bar{u}} = \left( \partial_{z} - \frac{\partial_z h_1}{4h_1} M \right) \phi_{\bar{u}} = 0,
$$
and so $\phi_{\bar{u}} = h_1^{M/4} \phi(\bar{z})$. Including also the $w$ direction we obtain
\be
\phi_{\bar{u}} = h_1^{M/4} h_2^{N/4} \phi(\bar{z}, \bar{w}),
\ee
where $\phi(\bar{z},\bar{w})$ is antiholomorphic.
This field has $R$-charge 2 in contrast to the
fermions which had $R$-charge 1. So when we twist the theory we need to modify the equations of motion by
two units of flux, $(M, N) \to (M-2, N-2)$,  rather than the single unit of flux that applied for the fermions. The fact that
$M \to M-2$ rather than $M \to M+2$ is by analogy with the twist for $\psi_3$ which is the fermionic partner of
$\phi$.

The Higgs field whose vev separates different brane stacks is only charged due to the twisting (as it is valued in the Cartan it is uncharged under any gauge flux
that is turned on). So in fact the twisting gives the sole `gauge' contribution to the wavefunction, and we have
\bea
\label{eqhiggsvev}
\phi_{\bar{u}} & = & h_1^{-1/2} h_2^{-1/2} \phi(\bar{z}, \bar{w}), \nonumber \\
\phi_{u} & = & h_1^{-1/2} h_2^{-1/2} \phi(z, w).
\eea
In order to use this vev to determine the twisted fermionic equations of motion we need to compute
$
-i \Gamma^r \left[ \phi_r, \lambda \right].
$
Now,
$$
\Gamma^u = 2 \left( \begin{array}{cc} \mbb{I} & 0 \\ 0 & - \mbb{I} \end{array} \right) \otimes
\left( \begin{array}{cc} \mbb{I} & 0 \\ 0 & - \mbb{I} \end{array} \right) \otimes \left( \begin{array}{cc} 0 & 1 \\ 0 & 0
\end{array} \right), \qquad
\Gamma^{\bar{u}} = 2 \left( \begin{array}{cc} \mbb{I} & 0 \\ 0 & - \mbb{I} \end{array} \right) \otimes
\left( \begin{array}{cc} \mbb{I} & 0 \\ 0 & - \mbb{I} \end{array} \right) \otimes \left( \begin{array}{cc} 0 & 0 \\ 1 & 0
\end{array} \right).
$$
With
$$
\lambda_1 = \left( \begin{array}{c} \psi_1 \\ 0 \end{array} \right) \otimes \left( \begin{array}{c} 1 \\ 0 \end{array}
\right), \qquad \lambda_2 = \left( \begin{array}{c} 0 \\ \psi_2 \end{array} \right) \otimes \left( \begin{array}{c} 0 \\ 1 \end{array}
\right),
$$
we therefore obtain
\be
i (\Gamma^u \phi_u + \Gamma^{\bar{u}} \phi_{\bar{u}}) (\lambda_1 + \lambda_2)
= 2 i \phi_u  \left( \begin{array}{c} 0 \\ \psi_2 \end{array} \right) \otimes
\left( \begin{array}{c} 1 \\ 0 \end{array} \right) - 2i \phi_{\bar{u}} \left( \begin{array}{c}
\psi_1 \\ 0 \end{array} \right) \otimes \left( \begin{array}{c} 0 \\ 1 \end{array} \right).
\ee
The equations of motion $\gamma^M D_M \lambda - i \Gamma^r [\phi_r, \lambda] = 0$ can also be written using
(\ref{abcd}) as
\bea
\label{first}
\mc{D}_{+} \psi_1 + 2i \phi_u \psi_2 & = & 0, \\
\label{second}
\mc{D}_{-} \psi_2 - 2i \phi_{\bar{u}} \psi_1 & = & 0.
\eea
Writing out (\ref{first}) and (\ref{second}) gives
\be
\label{first2}
\left( \begin{array}{cc}
\frac{i}{h_1^{\half}} \left( \partial_{\bar{z}} + \frac{\partial_{\bar{z}} h_1}{4h_1}(2+M) \right) & \frac{1}{h_2^{\half}} \left(
\partial_{\bar{w}} + \frac{\partial_{\bar{w}} h_2}{4h_2}(2+N) \right)  \\
\frac{1}{h_2^{\half}} \left( \partial_w - N\frac{\partial_w h_2}{4h_2} \right) & \frac{i}{h_1^{\half}} \left( \partial_z
- M\frac{ \partial_z h_1}{4h_1} \right) \end{array} \right) \left( \begin{array}{c} \psi_{1,1} \\ \psi_{1,2} \end{array}
\right) = -2i \phi_u \left( \begin{array}{c} \psi_{2,1} \\ \psi_{2,2} \end{array} \right).
\ee
\be
\label{second2}
\left( \begin{array}{cc}  \frac{-i}{h_1^{\half}} \left( \partial_z + \frac{\partial_z h_1}{4h_1}(2-M) \right) &
\frac{1}{h_2^{\half}}\left( \partial_{\bar{w}} + N\frac{\partial_{\bar{w}} h_2}{4 h_2} \right) \\
\frac{1}{h_2^{\half}} \left( \partial_w + \frac{\partial_w h_2}{4h_2} (2-N) \right) & \frac{-i}{h_1^{\half}} \left(
\partial_{\bar{z}} + M\frac{\partial_{\bar{z}} h_1}{4 h_1} \right) \end{array} \right) \left( \begin{array}{c} \psi_{2,1} \\
\psi_{2,2} \end{array} \right) = + 2i \phi_{\bar{u}} \left( \begin{array}{c} \psi_{1,1} \\ \psi_{1,2} \end{array} \right).
\ee
As we do not want to excite the gaugino we put $\psi_{2,2} = 0$. The second set of equations (\ref{second2}) then gives
\bea
\frac{-i}{h_1^{\half}} \left( \partial_z + \frac{\partial_z h_1}{4h_1} (2-M) \right) \psi_{2,1}
& = &  2 i \phi_{\bar{u}} \psi_{1,1}, \nonumber \\
\frac{1}{h_2^{\half}} \left( \partial_w + \frac{\partial_w h_2}{4h_2} (2-N)\right) \psi_{2,1} & = &
2i \phi_{\bar{u}} \psi_{1,2}.
\eea
Let us rewrite these equations as
\bea
\label{second31}
\mc{D}_z \left( h_1^{\half} h_2^{\half} \psi_{2,1} \right) = \left(\partial_z - \frac{M}{4h_1} \partial_z h_1 \right)(h_1^{\half} h_2^{\half} \psi_{2,1})
& = & 2 \phi_{\bar{u}} h_1^{\half} h_2^{\half} \left( -h_1^{\half} \psi_{1,1} \right),\\
\label{second32}
\mc{D}_w \left( h_1^{\half} h_2^{\half} \psi_{2,1} \right) = (\partial_w - \frac{N}{4h_2} \partial_w h_2 ) \left( h_1^{\half} h_2^{\half} \psi_{2,1} \right)
& = & 2 \phi_{\bar{u}} h_1^{\half} h_2^{\half} \left( i h_2^{\half} \psi_{1,2} \right).
\eea

Using $\psi_{2,2} = 0$ in (\ref{first2}), and also (\ref{second31}) and (\ref{second32}), we obtain
\bea
\label{eqeqc}
\Bigg[ \mc{D}_w \left( \frac{1}{\phi_{\bar{u}} h_1^{\half} h_2^{\half}} \mc{D}_z \right) - \mc{D}_z
\left( \frac{1}{\phi_{\bar{u}} h_1^{\half} h_2^{\half}} \mc{D}_w \right) \Bigg] \left( h_1^{\half} h_2^{\half}
\psi_{2,1} \right) = 0.
\eea
Note (\ref{eqeqc}) is automatically satisfied provided $\phi_{\bar{u}} h_1^{\half} h_2^{\half}$ is
an anti-holomorphic function of $z$ and $w$, as indeed occurs using (\ref{eqhiggsvev}). The commutator
$[D_z, D_w]$ gives a factor of flux $F_{zw}$ which vanishes for the factorised metric form used.

The upper equation of (\ref{first2}) likewise gives
\be
\label{eqeqd}
h_2 \mc{D}_{\bar{z}} \left( -h_1^{\half} \psi_{1,1} \right) + h_1 \mc{D}_{\bar{w}} \left( i h_2^{\half} \psi_{1,2} \right)
= 2 \left( h_1^{\half} h_2^{\half} \phi_u \right) \left( h_1^{\half} h_2^{\half} \psi_{2,1} \right).
\ee
Putting $\phi_{\bar{u}} = h_1^{-\half} h_2^{-\half} \frac{\bar{\phi}(\bar{z}, \bar{w})}{2}$ as in (\ref{eqhiggsvev})
(and introducing a factor of 2 for convenience), equations (\ref{eqeqd}) (\ref{second31}), (\ref{second32}) and
 (\ref{eqeqc}) become
 \bea
 h_2 \mc{D}_{\bar{z}} \left( -h_1^{\half} \psi_{1,1} \right) + h_1 \mc{D}_{\bar{w}} \left( i h_2^{\half} \psi_{1,2} \right)
 & = & \phi(z, w) \left( h_1^{\half} h_2^{\half} \psi_{2,1} \right). \label{aaa1} \\
 \mc{D}_z \left( h_1^{\half} h_2^{\half} \psi_{2,1} \right) = \bar{\phi}(\bar{z}, \bar{w}) \left( -h_1^{\half} \psi_{1,1} \right), \label{aaa2} \\
 \mc{D}_w \left( h_1^{\half} h_2^{\half} \psi_{2,1} \right) = \bar{\phi}(\bar{z}, \bar{w}) \left( i h_2^{\half} \psi_{1,2} \right), \label{aaa3} \\
 \frac{1}{\bar{\phi}(\bar{z}, \bar{w})} \left( \mc{D}_w \mc{D}_z - \mc{D}_z \mc{D}_w \right) \left( h_1^{\half} h_2^{\half} \psi_{2,1} \right)
 = 0. \label{aaa4}
\eea

Equations (\ref{aaa1}) to (\ref{aaa4}) can be identified with those coming from the topological theory of \cite{Beasley:2008dc}.
In fact, these equations are precisely identical (up to a complex conjugation which comes from choice of conventions) with
equations (\ref{preqofmot1}) to (\ref{preqofmot3}). The dictionary to map between the two formulations, namely direct reduction
of Super Yang-Mills and the topologically twisted theory, is given by
$$
- h_1^{\half} \psi_{1,1} \vert_{SYM} \to \psi_z \vert_{twist}, \qquad
i h_2^{\half} \psi_{1,2} \vert_{SYM} \to \psi_w \vert_{twist}, \qquad
h_1^{\half} h_2^{\half} \psi_{2,1} \vert_{SYM} \to \chi^{*} \vert_{twist},
$$
$$
h_1^{\half} h_2^{\half} \phi \vert_{SYM} \to \phi \vert_{twist}, \qquad
A \vert_{SYM} \to i A \vert_{twist}, \qquad \psi_{2,2} \vert_{SYM} \to A_{\mu} \vert_{twist}.
$$
We can verify these identifications by applying them to the normalisation of
 the kinetic terms. The normalisation prefactor for the kinetic terms is given
in the 8d SYM theory by
$$
\int_{\Sigma} d^4 y \, \sqrt{g} \psi^{\dagger} \psi,
$$
applying to all fermions (we shall tend to use $y$ for internal integrals over $S$ and $x$ for integrals over 8d space or
4d non-compact space). 
 Using the above dictionary it is easy to see that the kinetic terms are
\bea
\int d^4 y \, \sqrt{g} \psi_{1,1}^{*} \psi_{1,1} & \to & \int d^4 y \, h_2 \psi_{\bar{z}} \psi_z, \\
\int d^4 y \, \sqrt{g} \psi_{1,2}^{*} \psi_{1,2} & \to & \int d^4 y \, h_1 \psi_{\bar{w}} \psi_w, \\
\int d^4 y \, \sqrt{g} \psi_{2,1}^{*} \psi_{2,1} & \to & \int d^4 y \, \chi^{*} \chi.
\eea
The terms on the right-hand side are precisely those emerging from the kinetic terms $\int \omega \wedge A \wedge \bar{A}$
and $\int \chi \wedge \bar{\chi}$ in the Lagrangian in the appendix of \cite{Beasley:2008dc}.
In this way we see that the different kinetic terms
for matter in the topologically twisted theory
all arise from the same term $\int \sqrt{g} \psi^{\dagger} \psi$
in the super Yang-Mills theory.

\subsection{Yukawa Couplings}

The Yukawa couplings of the topologically twisted theory can be related to
Yang-Mills theory in the same way the equations of motion can.
 In dimensional reduction, all Yukawa couplings arise from the trilinear
 interactions in the Yang-Mills Lagrangian.
 This section will in part follow  \cite{08070789}, but will extend and improve the
 discussion from that paper.

 In the super Yang-Mills theory the Yukawa couplings all
 descend from the terms (dropping overall constants)
 \be
 \int d^8 x \sqrt{g} \, \Bigg[ \hbox{Tr} \left(\bar{\lambda}^I \Gamma^M A^J_M \lambda^K \right) + \hbox{Tr} \left(\bar{\lambda}^I
 \Gamma^r \phi^J_r \lambda^K \right) \Bigg].
 \ee
 Here $I,J,K$ are family indices.
Both these terms contribute to Yukawa couplings and we consider them separately. Let us start with the second case,
where the bosonic field comes from the transverse scalar $\phi^J$. In this case
\be
\Gamma^0 \Gamma^u = \left( \begin{array}{cc} 0 & \mbb{I} \\ \mbb{I} & 0 \end{array} \right)
\otimes \left( \begin{array}{cc} \mbb{I} & 0 \\ 0 & -\mbb{I} \end{array} \right) \otimes \left( \tau^u \right),
\ee
which gives a chirality flip in both the $0,1,2,3$ and $8,9$ directions.
To obtain a non-vanishing integral we therefore need
$\lambda_I$ and $\lambda_K$ to be either both of the form $(\lambda_1 + \lambda_4)$ or both of the form
$(\lambda_2 + \lambda_3)$.

We first assume the form $\lambda_I, \lambda_K = (\lambda_1 + \lambda_4)_{I,K}$, when the total Yukawa interaction is
\bea
\mc{L}_{YUK} & = & \int \sqrt{g} d^{8} x \, \, \left(
\lambda_{4,I}^{\dagger} \Gamma^0 \Gamma^M A_{M,J} \lambda_{1,K} +
\lambda_{1,I}^{\dagger} \Gamma^0 \Gamma^M A_{M,J} \lambda_{4,K} \right) \nonumber \\
& = &
\int \sqrt{g} d^{8} x \, \,
\left( \xi_{4I}^{\dagger}(x) \xi_{1,K}(x) \right) \left( \psi_{4,I}^{\dagger} (y) \psi_{1,K}(y) \right)
\begin{array}{c} \left( \begin{array}{cc} 0 & \theta_{4I}^{\dagger}(z) \end{array} \right) \\ \ph{0} \end{array}
\Bigg( \tau^a \phi_{aJ} \Bigg) \left( \begin{array}{c} \theta_{1K} \\ 0 \end{array} \right)
+ \nonumber \\
& & \left( \xi_{1I}^{\dagger}(x) \xi_{4,K}(x) \right) \left( \psi_{1,I}^{\dagger} (y) \psi_{4,K}(y) \right)
\begin{array}{c} \left( \begin{array}{cc} 0 & \theta_{1I}^{\dagger}(z) \end{array} \right) \\ \ph{0} \end{array}
\Bigg( \tau^a \phi_{aJ} \Bigg) \left( \begin{array}{c} \theta_{4K} \\ 0 \end{array} \right).
\eea
Using the relations (\ref{Majrelats}), we can put all expressions in terms of $\lambda_1$ alone, eliminating all
$\lambda_4$ dependence. As $\lambda_4$ corresponds to the CPT conjugate of $\lambda_1$, by
working only with $\lambda_1$ we are in effect working only with left-handed spinors, treating these as fundamental
as in the usual approach to constructing supersymmetric Lagrangians.
We also take $(\begin{array}{cc} \theta_1(z) & 0 \end{array}) = (\begin{array}{cc} 1 & 0 \end{array})$ as the transverse
direction is trivial. We obtain
\bea
\mc{L}_{YUK} & = &
2 \int
d^4 x \left( \xi_{1I}^T(x) \sigma_y \xi_{1K}(x) \phi_{\bar{u},J}(x) \right)
\int \sqrt{g} d^4 y \left( \psi_{1I}^T(y) \sigma_y \psi_{1K}(y) \right)
\phi_{\bar{u},J}(y) \quad + \nonumber \\
& & 2 \int d^4 x \left( \xi_{1I}^{\dagger}(x) \sigma_y \xi^{*}_{1K}(x) \phi_{u,J}(x) \right)
\int \sqrt{g} d^4 y \left( \psi_{1I}^{\dagger}(y) \sigma_y \psi_{1K}^{*}(y)
\right) \phi_{u,J}(y)  \\
& = & 2 \int d^4 x \left( \xi_{1I}^T(x) \sigma_y \xi_{1K}(x) \phi_{\bar{u},J}(x) \right)
\int \sqrt{g} d^4 y \left( \psi_{1I}^T(y) \sigma_y \psi_{1K}(y) \right)
\phi_{\bar{u},J}(y) + \hbox{c.c} \nonumber
\label{yukky}
\eea
Here $\phi_{u(\bar{u})}(y) = \phi_8(y) +(-) i \phi_9(y)$ and we have a trivial (flat) metric in the non-compact directions.
Now,
\bea
\psi_{1,I}^T \sigma_y \psi_{1,K}(y) \phi_{\bar{u},J} & = & \begin{array}{c} \left( \begin{array}{cc} \psi^I_{1,1} & \psi^I_{1,2} \end{array} \right)
\\ \ph{0} \end{array}
\left( \begin{array}{cc} 0 & -i \\ i & 0 \end{array} \right)  \left( \begin{array}{c} \psi^K_{1,1} \\ \psi^K_{1,2} \end{array} \right) \nonumber \\
& = & i ( \psi_{1,2}^I \psi_{1,1}^K - \psi_{1,1}^I \psi_{1,2}^K ) \phi_{\bar{u},J}.
\eea
The total interaction is therefore
\be
\label{YukA}
2i \int d^4 x \left( \xi_{1,I}^T \sigma_y  \xi_{1,K} \phi_{\bar{u},J} \right) \int \sqrt{g} d^4 y \,  \left( \psi_{1,2}^I \psi_{1,1}^K
- \psi_{1,1}^I \psi_{1,2}^K \right) \phi_{\bar{u},J} + \hbox{c.c}
\ee

There is also another possible term, where we let $\lambda^I = (\lambda_2 + \lambda_3)$ and $\lambda^K = (\lambda_2 + \lambda_3)$.
 However these interactions do not give rise to Yukawa couplings. These couplings always involve a contribution from
 $\psi_{2,2}$. The mode $\psi_{2,2}$ corresponds to the gaugino that partners
the 4d gauge boson. Such interactions are therefore gauge rather than Yukawa couplings.

So instead we move on to consider terms coming from the interaction
$$
\hbox{Tr}( \bar{\lambda}^I \Gamma^M [A_M^J, \lambda^K]),
$$
where we now draw the scalar part of the Yukawa coupling from the $A_M$ vector field.
In this case
\be
\Gamma^0 \Gamma^M = \left( \begin{array}{cc} 0 & \mbb{I} \\ \mbb{I} & 0 \end{array} \right)
\otimes \Bigg( \, \, \, \, \tilde{\gamma}^m \, \, \, \, \Bigg) \otimes \mbb{I},
\ee
which generates a chirality flip in both the $0,1,2,3$ and $4,5,6,7$ directions.
In this case to obtain a non-vanishing
integral $\lambda_I$ and $\lambda_K$ must take different forms. There are two options:
we first consider the case $\lambda_K = (\lambda_1 + \lambda_4)$, and $\lambda_I =
(\lambda_2 + \lambda_3)$, and subsequently analyse the case $\lambda_K = (\lambda_2 + \lambda_3)$,
$\lambda_I = (\lambda_1 + \lambda_4)$.

For the first case the total Yukawa interaction is
\bea
\label{firstc}
\mc{L}_{YUK} & = & \int  d^{8} x \sqrt{g} \, \left(
\lambda_{3,I}^{\dagger} \Gamma^0 \Gamma^M A_{M,J} \lambda_{1,K} +
\lambda_{2,I}^{\dagger} \Gamma^0 \Gamma^M A_{M,J} \lambda_{4,K} \right) \nonumber \\
& = & \int d^{8} x \sqrt{g} \quad
\Bigg[ \left( \xi_{3,I}^{\dagger}(x) \xi_{1,K}(x) \right)
\begin{array}{c} \left(  \begin{array}{cc} 0 & \psi_{3,I}^{\dagger} (y) \end{array} \right)  \\ \ph{0} \end{array}
\Bigg( \tilde{\gamma}^M A_{MJ} \Bigg) \left( \begin{array}{c} \psi_{1,K}(y) \\ 0 \end{array} \right)
\left( \theta_{3,I}^{\dagger}(z) \theta_{1,K} \right)
+ \nonumber \\
& & \left( \xi_{2,I}^{\dagger}(x) \xi_{4,K}(x) \right)
\begin{array}{c}
\left(  \begin{array}{cc} 0 & \psi_{2,I}^{\dagger} (y) \end{array} \right) \\ \ph{0} \end{array}
\Bigg( \tilde{\gamma}^M A_{MJ} \Bigg) \left( \begin{array}{c} \psi_{4,K}(y) \\ 0 \end{array} \right)
\left( \theta_{2,I}^{\dagger}(z) \theta_{4,K} \right) \Bigg].
\eea
We again use the relations (\ref{Majrelats}) to write everything in terms of $\lambda_1$ and
$\lambda_2$, and take $(\begin{array}{cc} \theta_1(z) & 0 \end{array}) = (\begin{array}{cc} 1 & 0 \end{array})$.
The Yukawa interactions then become
\bea
\mc{L}_{YUK} & = & -\int
d^4 x \left( \xi_{2I}^T(x) \sigma_y \xi_{1K}(x) A_{J}(x) \right)
\int d^4 y \sqrt{g}
\begin{array}{c} \left( \begin{array}{cc} 0 & \psi_{2,I}^{T} \sigma_y \end{array} \right) \\ \ph{0} \end{array}
\Bigg( \tilde{\gamma}^M A_{M,J} \Bigg) \left( \begin{array}{c} \psi_{1,K}(y) \\ 0 \end{array} \right)
\nonumber \\
& + & \int d^4 x \begin{array}{c} \left( \xi_{2I}^{\dagger}(x) \sigma_y \xi^{*}_{1K}(x) A_{J}(x) \right) \\ \ph{0} \end{array}
\int d^4 y \sqrt{g}
\begin{array}{c} \left( \begin{array}{cc} 0 & \psi_{2,I}^{\dagger} \end{array} \right) \\ \ph{0} \end{array}
\Bigg( \tilde{\gamma}^M A_{MJ} \Bigg) \left( \begin{array}{c}\sigma_y \psi^{*}_{1,K}(y) \\ 0 \end{array} \right). \nonumber
\eea
Using the gamma matrices we can now write
\be
\tilde{\gamma}^M A_M = 2 \left( \begin{array}{cccc} 0 & 0 & -i A_{z} & A_{\bar{w}} \\
0 & 0 & A_{w} & -iA_{\bar{z}} \\ iA_{\bar{z}} & A_{\bar{w}} & 0 & 0 \\ A_{w} & i A_{z} & 0 & 0 \end{array}
\right),
\ee
with
$\psi_2^T \sigma_y = i \left( \begin{array}{cc} \psi_{2,2} & - \psi_{2,1} \end{array} \right)$.
So
\bea
\begin{array}{c} \left( \begin{array}{cc} 0 & \psi_{2,I}^{T} \sigma_y \end{array} \right) \\ \ph{0} \end{array}
\Bigg( \tilde{\gamma}^M A_{M,J} \Bigg) \left( \begin{array}{c} \psi_{1,K}(y) \\ 0 \end{array} \right)
& = &
2i \begin{array}{c} \left( \begin{array}{cccc} 0 & 0 & \psi^I_{2,2} & -\psi^I_{2,1} \end{array} \right) \\ \ph{0}
\\ \ph{0} \\ \ph{0} \end{array}
\left( \begin{array}{cccc} 0 & 0 & -i A_{z}^J & A^J_{\bar{w}} \\
0 & 0 & A^J_{w} & -iA^J_{\bar{z}} \\ iA^J_{\bar{z}} & A^J_{\bar{w}} & 0 & 0 \\ A^J_{w} & i A^J_{z} & 0 & 0 \end{array}
\right)
 \left( \begin{array}{c}
\psi^K_{1,1} \\ \psi^K_{1,2} \\ 0 \\ 0 \end{array} \right) \nonumber \\
& = & 2i \left( - \psi_{2,1}^I \left( A_{\bar{w}}^J \psi_{1,1}^k + i A_{\bar{z}}^J \psi_{1,2}^K \right)
+ \psi_{2,2}^I \left( i \psi_{1,1}^K A_z^J + A_w^J \psi_{1,2}^K \right) \right). \nonumber
\eea
We also need to evaluate the conjugate expression
\bea
\begin{array}{c}
\left( \begin{array}{cc} 0 & \psi_{2,I}^{\dagger} \end{array} \right) \\ \ph{0} \end{array}
\Bigg( \tilde{\gamma}^M A_{MJ} \Bigg) \left( \begin{array}{c}\sigma_y \psi^{*}_{1,K}(y) \\ 0 \end{array} \right)
& = &
2i \left( \begin{array}{cccc} 0 & 0 & \psi_{2,1}^{I,*} & \psi_{2,2}^{I,*} \end{array} \right)
\left( \begin{array}{cccc} 0 & 0 & -i A_{z}^J & A^J_{\bar{w}} \\
0 & 0 & A_{w} & -iA_{\bar{z}} \\ iA_{\bar{z}}^J & A^J_{\bar{w}} & 0 & 0 \\ A^J_{w} & i A^J_{z} & 0 & 0 \end{array}
\right)
\left( \begin{array}{c} -\psi_{1,2}^{K,*} \\ \psi_{1,1}^{K,*} \\ 0 \\ 0 \end{array} \right) \nonumber \\
& = & 2i \left( -\psi_{2,1}^{I,*} \left( - A_{\bar{w}}^J \psi_{1,1}^{K,*} + i A_{\bar{z}}^J \psi_{1,2}^K  \right)
+ \psi_{2,2}^{I,*} \left( i \psi_{1,1}^{K,*} A_{z}^J - A_{w}^J \psi_{1,2}^{K,*} \right) \right). \nonumber
\eea
Combining these we evaluate (\ref{firstc}) as
\be
\label{YukB}
2 \int d^{8} x \sqrt{g} \, \left( \xi_2^{I,T} \sigma_y \xi_1^K  \right) \left( \psi_{2,1}^I
\left( - A_{z}^J \psi_{1,2}^K + i A_{w}^J \psi_{1,1}^k  \right)
+ \psi_{2,2}^I \left( -i A_{\bar{w}}^J \psi_{1,2}^K + \psi_{1,1}^K A_{\bar{z}}^J  \right) \right) + \hbox{c.c}
\ee
The first part of this expression we can interpret as a Yukawa interaction. The second part (involving $\psi_{2,2}$)
should be interpreted as a gauge interaction as it involves the 4-dimensional gaugino.

Finally we examine the case of $\lambda^I = \lambda_1 + \lambda_4$ and $\lambda^K = \lambda_2 + \lambda_3$. This gives
\bea
\label{yukyuk}
\mc{L}_{YUK} & = &
\int d^{8} x \sqrt{g} \,
\left( \lambda_{4,I}^{\dagger} \Gamma^0 \Gamma^M A_{M,J} \lambda_{2,K} +
\lambda_{1,I}^{\dagger} \Gamma^0 \Gamma^M A_{M,J} \lambda_{3,K} \right), \\
& = & \int
d^4 x \left( \xi_{1I}^T(x) \sigma_y \xi_{2K}(x) \right) \int d^4 y \sqrt{g} \,
\begin{array}{c} \left( \begin{array}{cc} \psi_{1,I}^{T} \sigma_y & 0 \end{array} \right) \\ \ph{0} \end{array}
\Bigg( \tilde{\gamma}^M A_{MJ} \Bigg) \left( \begin{array}{c} 0 \\ \psi_{2,K}(y) \end{array} \right)
 \nonumber \\
& & -\int d^4 x \left( \xi_{1I}^{\dagger}(x) \sigma_y \xi^{*}_{2K}(x) \right) \int d^4 y \sqrt{g} \,
\begin{array}{c} \left( \begin{array}{cc} \psi_{1,I}^{\dagger} & 0 \end{array} \right) \\ \ph{0} \end{array}
\Bigg( \tilde{\gamma}^M A_{M,J} \Bigg) \left( \begin{array}{c} 0 \\ \sigma_y \psi^{*}_{2,K}(y)  \end{array} \right). \nonumber
\eea
We find
\bea
\begin{array}{c} \left( \begin{array}{cc} \psi_{1,I}^{T} \sigma_y & 0 \end{array} \right) \\ \ph{0} \end{array} \Bigg( \tilde{\gamma}^M A_{MJ} \Bigg) \left( \begin{array}{c} 0 \\ \psi_{2,K}(y) \end{array} \right) & = &
2i \left( \psi_{2,1}^K ( -i A^J_{z} \psi_{1,2}^I  - \psi_{1,1}^I A_{w}^J ) + \psi_{2,2}^K ( A_{\bar{w}} \psi_{1,2}^I
+ i A_{\bar{z}}^J \psi_{1,1}^I ) \right) \nonumber \\
- \begin{array}{c} \left( \begin{array}{cc} \psi_{1,I}^{\dagger} \sigma_y & 0 \end{array} \right) \\ \ph{0} \end{array}
\Bigg( \tilde{\gamma}^M A_{M,J} \Bigg) \left( \begin{array}{c} 0 \\ \sigma_y \psi^{*}_{2,K}(y)  \end{array} \right) & = &
-2i \left( \psi_{2,1}^{K,*} \left( \psi_{1,1}^{I,*} A_{\bar{w}}^J - i \psi_{1,2}^{I,*} A_{\bar{z}}^J \right)
+ \psi_{2,2}^{*} \left( i \psi_{1,1}^{I,*} A_{z}^J - \psi_{1,2}^{I,*} A_{w}^J \right) \right). \nonumber
\eea
We evaluate (\ref{yukyuk}) as
\be
\label{YukC}
2 \int d^{8}x \sqrt{g} \, \left( \xi_{1I}^T(x) \sigma_y \xi_{2K}(x) \right)
\left( \psi_{2,1}^K ( A^J_{z} \psi_{1,2}^I  - i \psi_{1,1}^I A_{w}^J ) +
\psi_{2,2}^K ( i A_{\bar{w}} \psi_{1,2}^I
- A_{\bar{z}}^J \psi_{1,1}^I ) \right) + \hbox{c.c}.
\ee
As for (\ref{YukA}), the first part of this expression should be interpreted as a Yukawa interaction and the second part
as a gauge interaction.

Let us gather together the three contributions to Yukawa interactions from (\ref{YukA}), (\ref{YukB}) and (\ref{YukC}). Put together, these give
\bea
& & 2 \int d^8 x \sqrt{g} \, \Bigg( \left( \xi_1^{T,I} \sigma_y \xi_1^K \right) \left[ i \left( \psi_{1,2}^I \psi_{1,1}^K - \psi_{1,1}^I \psi_{1,2}^K \right)
\phi_{\bar{u}}^J \right] + \\
& & \left( \xi_2^{T,I} \sigma_y \xi_1^K \right) \left[ \psi_{2,1}^I \left( -A_{z}^J \psi_{1,2}^K + i A_{w}^J \psi_{1,1}^K \right) \right] +
\left( \xi_1^{T,I} \sigma_y \xi_2^K \right) \left[ \psi_{2,1}^K
\left( A_{z} \psi_{1,2}^I - i \psi_{1,1}^I A_{w}^J \right) \right] \Bigg) + \hbox{c.c} \nonumber
\eea
There are a total of six different interactions contributing to Yukawa couplings. Using the above dictionary,
these Yukawa couplings can be seen to all descend from
$A^I \wedge A^J \wedge \Phi^k$ on cyclically permuting indices.

As both the equations of motion, kinetic terms and Yukawa couplings are precisely the same from both dimensional
reduction of super Yang-Mills and via the twisted theory, we are free to compute with either. One advantage of the Yang-Mills
formalism is that it connects with a set of intuitions that are not as manifestly obvious in the language of the twisted theory.

%%%%%%%%%%%%%%%%%%%%%%%%%%%%%%%%%%%%%%%%%%%%%%%%%%%%%%%%%%%%%%%%%%%%%%%%%%%%%
\subsection{Holomorphy Constraints on the Effective Theory}
\label{sec:modhol}
%%%%%%%%%%%%%%%%%%%%%%%%%%%%%%%%%%%%%%%%%%%%%%%%%%%%%%%%%%%%%%%%%%%%%%%%%%%%%

Classical Yang-Mills theory has simple scaling properties and so this suggests that any theory which can be
related to Yang-Mills simply by field redefinitions should also possess a similar property, namely that
rescalings of the metric can have only a very limited effect on the physical properties of the theory.

Let us recall the equations of motion for the topological theory,
\bea
\label{eom}
\omega \wedge \partial_A \psi + \frac{i}{2} \left[ \bar{\phi}, \chi \right] & = & 0, \nonumber \\
\bar{\partial}_A \chi - \left[ \phi, \psi \right] & = & 0.
\eea
Here $(\phi, \chi)$ are grouped inside the same superfield $\Phi$ as are $(A, \psi)$.
The Yukawa couplings are given by expanding $\int A \wedge A \wedge \Phi$ and the kinetic terms
come from the terms
$$
\int d^8 x \left( -2 \omega \wedge D_{\mu} \bar{\psi} \wedge \bar{\sigma}^\mu \psi \right),
\qquad \int d^8 x \left( D_{\mu} \chi \wedge \sigma^{\mu} \bar{\chi} \right).
$$
$\phi$ can attain an arbitrary holomorphic vev within the Cartan subalgebra. We suppose
$$
\langle \phi \rangle = f(z_1, z_2) t^1 + g(z_1, z_2) t^2
$$
with the matter curves located at $f=0$, $g=0$ and $f+g = 0$.

There is one important and simple set of solutions to (\ref{eom}),
given by the rescalings
\bea
\label{rescaling}
\phi & \to & \lambda \phi, \nonumber \\
\chi & \to & \lambda \chi, \nonumber \\
\omega & \to & \lambda^2 \chi, \nonumber \\
(A, \psi) & \to & (A, \psi).
\eea
It is manifest that given a set of solutions to (\ref{eom}) this rescaling
provides a new set of solutions. The rescalings of (\ref{rescaling}) correspond to
the overall metric rescaling $g_{ij} \to \lambda^2 g_{ij}$. To see this, note first that this
rescales the K\"ahler form as in (\ref{rescaling}). The scaling of $\phi$ is most easily seen
by considering the type IIB case, where $\phi$ can be interpreted as the transverse separation of
the brane stacks. Such a transverse separation grows linearly with the overall length scale
consistent with (\ref{rescaling}). $\chi$ belongs in the same supermultiplet as $\phi$ and so scales in the same fashion.

$A$ does not rescale, and so the unnormalised Yukawa couplings scale as
$$
\int A \wedge A \wedge \Phi \longrightarrow \lambda \int A \wedge A \wedge \Phi,
$$
as every contributing term has one factor of $\chi$.
The kinetic terms scale as
$$
\int \omega \wedge D_{\mu} \bar{\psi} \wedge \sigma^{\mu} \psi \to
\lambda^2 \int \omega \wedge D_{\mu} \bar{\psi} \wedge \sigma^{\mu} \psi, \quad
\int D_{\mu} \chi \wedge \sigma^{\mu} \bar{\chi} \to
\lambda^2 \int D_{\mu} \chi \wedge \sigma^{\mu} \bar{\chi},
$$
and so all kinetic terms scale as $Z \to \lambda^2 Z$.

The normalised (physical) Yukawa couplings therefore scale as
$$
\frac{ \int A^a \wedge A^b \wedge \Phi^c}{\sqrt{Z_a Z_b Z_c}} \to \frac{ \lambda \int A^a \wedge A^b \wedge \Phi^c}
{\sqrt{ (\lambda^2 Z_a) (\lambda^2 Z_b) (\lambda^2 Z_c)}} = \frac{1}{\lambda^2}\frac{ \int A^a \wedge A^b \wedge \Phi^c}{\sqrt{Z_a Z_b Z_c}}.
$$
The metric rescaling increases the size of the 4-cycle on which the 7-branes supporting the Yang-Mills theory live, and therefore
rescales $\alpha_{GUT}^{-1} \to \lambda^4 \alpha_{GUT}^{-1}$. Physical Yukawa couplings therefore behave
as $Y^{phys}_{\alpha \beta \gamma} \sim \alpha_{GUT}^{1/2}$, with $\alpha_{GUT}$ appearing as a universal prefactor.

This conclusion could have been anticipated based on holomorphy and the structure of 4-dimensional supergravity theories.
These provide very powerful general constraints on the structure of the effective action.
The gauge coupling $\alpha^{-1}$ appears in the action as the coefficient of the kinetic
term $F_{\mu \nu} F^{\mu \nu}$. In string theory the gauge coupling is dynamical, and so is part of a
K\"ahler modulus\footnote{We restrict to language appropriate to IIB/F-theory compactifications.}
$T_{GUT} = \frac{4 \pi}{g^2} + i a$, whose imaginary part is an axion carrying a perturbative shift symmetry.

As the imaginary part is axionic there is a perturbative shift symmetry $T \to T + i \epsilon$, which implies that
$T$ can make no perturbative appearance in the superpotential. Consequently the superpotential
Yukawa couplings cannot depend on $\alpha_{GUT}$, which only enters the Yukawa couplings via prefactors
derived from the K\"ahler potential.

The absence of $\alpha_{GUT}$ from the superpotential is a specific example of a general phenomenon: as all K\"ahler moduli
have shift symmetries which are exact within perturbation theory, the superpotential cannot depend on K\"ahler moduli.
However, from an effective field theory viewpoint the superpotential can have arbitrary dependence on complex structure moduli
and there is no reason not to expect a rank 3 Yukawa matrix.

%%%%%%%%%%%%%%%%%%%%%%%%%%%%%%%%%%%%%%%%%%%%%%%%%%%%%%%%%%%%%%%%%%%%%%%%%%%%%
\section{Wavefunctions and Yukawas with background fluxes}
\label{sec:yukcoup}
%%%%%%%%%%%%%%%%%%%%%%%%%%%%%%%%%%%%%%%%%%%%%%%%%%%%%%%%%%%%%%%%%%%%%%%%%%%%%

In this section we explicitly solve the equations of motion to obtain the matter wavefunctions in the presence of background flux. In section \ref{sec:fluxd} we solve the D-term equations which require a perturbed background metric and discuss the perturbative expansion used in finding the wavefunction solutions. Subsequently in sections \ref{sec:mattwav} and \ref{sec:higgswav} we present solutions to the equations of motion. In section \ref{sec:yukcoupfinal} we use the wavefunction profiles to calculate the resulting Yukawa couplings, finding no corrections to a rank-1 matrix.

%%%%%%%%%%%%%%%%%%%%%%%%%%%%%%%%%%%%%%%%%%%%%%%%%%%%%%%%%%%%%%%%%%%%%%%%%%%%%
\subsection{Flux solutions to D-terms}
\label{sec:fluxd}
%%%%%%%%%%%%%%%%%%%%%%%%%%%%%%%%%%%%%%%%%%%%%%%%%%%%%%%%%%%%%%%%%%%%%%%%%%%%%

We first set out the background flux configuration. We consider only Abelian flux and so there are three relevant generators for a rank 2 enhancement point. We denote these $Q_0$, $Q_1$ and $Q_2$, with $Q_0$ corresponding to the hypercharge flux. We denote the corresponding flux and gauge fields with the index $I$ such that
\be
A = \sum_I A_{I} = A_{0} + A_{1} + A_{2} \;.
\ee
The constraints on the flux from the equations of motion are
\be
F^{(2,0)} = F^{(0,2)} = 0 \;.
\ee
We also have the D-term
\be
\omega \wedge F^{(1,1)} = 0 \;.
\ee
In this paper we consider the factorised case with no $dz_1 \wedge d\bar{z}_{\bar{2}}$ terms
\be
F = F_{1\bar{1}} dz_1 \wedge d\bar{z}_{\bar{1}} + F_{2\bar{2}} dz_2 \wedge d\bar{z}_{\bar{2}} \;.
\ee
Apart from simplicity, cross terms in the flux do not affect the D-term equation for a metric ansatz (\ref{metricans}) and so we do not expect new effects regarding their compatibility with supersymmetry.
We take the expansion to first order as in \cite{Font:2009gq} and also define $\tilde{F}$ through
\bea
F_{1\bar{1}} &=& 2iM + 4i\left( \alpha_1 z_1 + \bar{\alpha}_1 \bar{z}_{\bar{1}} \right) \equiv 2iM + i\tilde{F}_{1\bar{1}} \;, \\
F_{2\bar{2}} &=& 2iN + 4i\left( \alpha_2 z_2 + \bar{\alpha}_2 \bar{z}_{\bar{2}} \right) \equiv 2iN + i\tilde{F}_{2\bar{2}} \;. \label{fluxans}
\eea
Here $M$, $N$, $\alpha_i$, $\tilde{F}$ all have a suppressed generator index $I$. Next we perform a gauge transformation as in \cite{Font:2009gq}
\bea
\hat{A} = A - d\Omega \;,
\eea
such that $\hat{A}_{\bar{1}}=\hat{A}_{\bar{2}}=0$. The corresponding potentials read
\bea
\hat{A}_1 &=& -2iM \bar{z}_{\bar{1}} - 2i\left( \bar{\alpha}_1 \bar{z}^2_{\bar{1}} + 2\alpha_1 z_1 \bar{z}_{\bar{1}} \right) \;, \nn \\
\hat{A}_2 &=& -2iN \bar{z}_{\bar{2}} - 2i\left( \bar{\alpha}_2 \bar{z}^2_{\bar{2}} + 2\alpha_2 z_2 \bar{z}_{\bar{2}} \right) \;.
\eea
Taking the Kahler form as
\be
\omega = \frac{i}{2}\left( 1 + f_1\right) dz_1 \wedge d\bar{z}_{\bar{1}} + \frac{i}{2}\left( 1 + f_2\right) dz_2 \wedge d\bar{z}_{\bar{2}} \;,
\ee
the D-term is solved by
\be
M = -N \;, \qquad f_1 = -\frac{\tilde{F}_{1\bar{1}}}{2N} \;, \qquad f_2 = \frac{\tilde{F}_{2\bar{2}}}{2N} \;. \label{Dterm}
\ee
This fixes the Kahler form, up to an overall constant rescaling.\footnote{Note that the Bianchi identity for the flux ensures the Kahler condition on the metric.} It is important to note that the D-terms require a non-trivial metric background. Note that $f_1$ and $f_2$ are $O(1)$ in the flux,
 which will imply that the leading metric-induced perturbations to the wavefunctions dominates over the flux-induced perturbations. In turn this will mean that the wavefunction solutions that we find will be different to those presented in \cite{Font:2009gq}.

 There are three copies of the equations (\ref{Dterm}) due to the generator index $I$. This then implies
\be
\frac{\alpha_{i,I}}{N_I} = \frac{\alpha_{i,J}}{N_J} \equiv n_i \;\;\;\;\;\; \forall \;I,J \;,
\ee
where we define $n_i$ through
\bea
f_1 &=& -2\left( n_1 z_1 + \bar{n}_{\bar{1}} \bar{z}_{\bar{1}} \right) \nn \;, \\
f_2 &=& 2\left( n_2 z_2 + \bar{n}_{\bar{2}} \bar{z}_{\bar{2}} \right) \;.
\eea
We now give a vev to $\varphi$ such that
\be
\left< \varphi \right> = \frac{z_1}{v} Q_1 + \frac{z_2}{v} Q_2 \;. \label{vevphi2}
\ee
The equations of motion to be solved then read
\bea
& &\left(1+ f_1\right) \left( \partial_2 - i q\cdot \hat{A}_{2} \right) \hat{\psi}_{\bar{2}} + \left(1+ f_2\right) \left( \partial_1 - i q\cdot \hat{A}_{1}  \right) \hat{\psi}_{\bar{1}} -  \frac{1}{v}\left( \bar{z}_1 q_1 + \bar{z}_2 q_2\right) \hat{\chi} = 0\;, \label{eqofmot1} \nn \\
& & \bar{\partial}_{\bar{1}} \hat{\chi} - \frac{1}{v}\left( z_1 q_1 + z_2 q_2\right) \hat{\psi}_{\bar{1}} = 0 \;, \label{eqofmot2} \nn \\
& & \bar{\partial}_{\bar{2}} \hat{\chi} - \frac{1}{v}\left( z_1 q_1 + z_2 q_2\right) \hat{\psi}_{\bar{2}} = 0 \;. \label{eqofmot3}
\eea
Here $q\cdot$ denotes contraction into the vector of charges $q = \left(q_0, q_1, q_2\right)$.

In order to solve for the wavefunctions we perform a perturbative expansion. In \cite{Heckman:2008qa} the relevant expansions were identified as the derivative expansion, which is an expansion in $v^{1/2}$, and the flux expansion, which is an expansion in $\alpha_i v^{3/2}$. Which one dominated depended on the size of the flux parameters $\alpha_i$ and their higher order analogues. In our case the appropriate expansions are in the metric deformations as these dominate over the flux contributions to the wavefunctions. To see this note that the flux contributions to the wavefunctions come in at $O(\alpha_i v^{3/2})$ \cite{Heckman:2008qa,Font:2009gq}, while using the solutions (\ref{2ndomat}) we see that the leading contribution from the metric expansion comes in at $O(n_i v^{1/2})$. This is leading in powers of $v$ but also note that $n_i \gg \alpha_i$ since $M \ll 1$. In the coming sections we solve the wavefunctions to second order in the `metric' expansion.\footnote{We only 
 calculate the contributions from the fluxes \ref{fluxans}, i.e. we do not consider higher order terms such as $\beta_i z_1^2$. For the first order solutions this does not matter since they are subleading, but they could possibly compete at second order. This is highly dependent on the choice for the parameters and so for simplicity we drop such terms.}

%%%%%%%%%%%%%%%%%%%%%%%%%%%%%%%%%%%%%%%%%%%%%%%%%%%%%%%%%%%%%%%%%%%%%%%%%%%%%
\subsection{Matter wavefunctions}
\label{sec:mattwav}
%%%%%%%%%%%%%%%%%%%%%%%%%%%%%%%%%%%%%%%%%%%%%%%%%%%%%%%%%%%%%%%%%%%%%%%%%%%%%

The matter curves correspond to the curves $z_1=0$ and $z_2=0$. By symmetry it is sufficient to solve for one of the curves which in this case we take to be the $z_1=0$ curve. In terms of the charges defined in section \ref{sec:fluxd} the curve corresponds to
\be
\left(q_0,q_1,q_2\right) = \left(q_0,1,0\right) \;.
\ee
Since we are only solving the equations of motion to ${\cal O}\left(z_i\right)={\cal O}\left(v^{1/2}\right)$ \footnote{Note that a wavefunction solution at ${\cal O}\left(z_i^r\right)$, for some $r$, solves the equations of motion to ${\cal O}\left(z_i^{r-1}\right)$.} we can drop the flux terms and so the equations of motion read
\bea
& &\left(1+ f_1\right) \partial_2 \psi_{\bar{2}} + \left(1+ f_2\right) \partial_1 \psi_{\bar{1}} - \frac{\bar{z}_1}{v} \chi = 0\;, \label{nofeqofmot1} \\
& & \bar{\partial}_{\bar{1}} \chi - \frac{z_1}{v} \psi_{\bar{1}} = 0 \;, \label{nofeqofmot2} \\
& & \bar{\partial}_{\bar{2}} \chi - \frac{z_1}{v} \psi_{\bar{2}} = 0 \;. \label{nofeqofmot3}
\eea
The equations (\ref{nofeqofmot2}) and (\ref{nofeqofmot3}) are trivially solved exactly for $\psi_{\bar{1}}$ and $\psi_{\bar{2}}$ in terms of $\chi$.

The perturbative solutions are based on the zeroth order solution, obtained by taking $f_1=f_2=0$, which we reproduce here for convenience
\be
\chi^{(0)} = g\left(z_2\right)e^{-\frac{|z_1|^2}{v}} \;. \label{zewa}
\ee
Here $g$ is a holomorphic function of $z_2$ which carries a suppressed generation index $g_a\left(z_2\right) = \left\{1,z_2,z_2^2\right\}$. The full set of equations are solved to ${\cal O}\left(z\right)$ by the ${\cal O}\left(z^2\right)$ solution
\bea
\chi &=& \chi^{(0)} \left\{ 1 + \frac{|z_1|^2}{v} \left[ \frac{f_2}{2} + \frac{1}{2}v\bar{n}_{\bar{2}} \left(\partial_2\mathrm{ln}g\right) - \frac{3f_2^2}{8} - \frac{5}{4}v|n_2|^2 - v\bar{n}_{\bar{2}}f_2\left(\partial_2\mathrm{ln}g\right) \right. \right.  \\
& & \left. - \frac{7}{8}v^2\bar{n}_{\bar{2}}^2 \left(\left(\partial_2\partial_2\mathrm{ln}g\right) + \left(\partial_2\mathrm{ln}g\right)^2 \right) \right] + \frac{|z_1|^4}{v^2} \left[\frac{f_2^2}{8} + \frac{1}{4}v|n_2|^2 + \frac{1}{4}v\bar{n}_{\bar{2}}f_2\left(\partial_2\mathrm{ln}g\right) \right. \nn \\
& & \left. \left. + \frac{1}{8}v^2\bar{n}_{\bar{2}}^2 \left( \left(\partial_2\partial_2\mathrm{ln}g\right) + \left(\partial_2\mathrm{ln}g\right)^2 \right) \right] + \frac{\bar{n}_2}{3}\left(\partial_2\mathrm{ln}g\right)f_1|z_1|^2 - \frac{2v}{3}z_1n_1\bar{n}_2\left(\partial_2\mathrm{ln}g\right)   \right\} \;. \nonumber \label{2ndomat}
\eea

The solution for the second matter curve $z_2=0$ is obtained from (\ref{2ndomat}) by the operations
\be
z_1 \leftrightarrow z_2 \;,\;\; n_1 \leftrightarrow -n_2 \;\;\left( \implies f_1 \leftrightarrow f_2 \right) \;.
\ee

%%%%%%%%%%%%%%%%%%%%%%%%%%%%%%%%%%%%%%%%%%%%%%%%%%%%%%%%%%%%%%%%%%%%%%%%%%%%%
\subsection{Higgs wavefunctions}
\label{sec:higgswav}
%%%%%%%%%%%%%%%%%%%%%%%%%%%%%%%%%%%%%%%%%%%%%%%%%%%%%%%%%%%%%%%%%%%%%%%%%%%%%

The Higgs curve is the curve $z_1+z_2=0$ which corresponds to charges
\be
\left(q_0,q_1,q_2\right) = \left(q_0,-1,-1\right) \;.
\ee
The order counting of the derivative expansion treats $z_1$ and $z_2$ on an equal footing and so applies unaltered to this solution. Again we solve the equations of motion to ${\cal O}\left(z\right)$ obtaining a solution for $\chi$ to ${\cal O}\left(z^2\right)$. The flux terms in the equations of motion only come in at higher orders and so can be dropped.

As suggested in \cite{Font:2009gq}, to find the solutions it is convenient to define
\bea
w &\equiv& z_1 + z_2 \;, \;\; u \equiv z_1 - z_2 \;, \nn \\
\psi_{\bar{w}} &\equiv& \frac12\left(\psi_{\bar{1}} + \psi_{\bar{2}} \right)\;, \;\; \psi_{\bar{u}} \equiv \frac12\left( \psi_{\bar{1}} - \psi_{\bar{2}} \right) \;, \nn \\
n_+ &\equiv& \frac{1}{2} \left(n_1+n_2\right) \;, \;\; n_- \equiv \frac{1}{2} \left(n_1-n_2\right) \;, \nn \\
f_+ &\equiv& -\frac12 \left( f_1 + f_2 \right) = n_+u + n_-w + \bar{n}_+\bar{u} +\bar{n}_- \bar{w} \;, \nn \\
f_- &\equiv& \frac12 \left( f_2 - f_1\right) = n_+w + n_-u + \bar{n}_+\bar{w} +\bar{n}_- \bar{u} \;.
\label{lalala}
\eea
With this change of variables the equations of motion read
\bea
& &\left(1 - f_+\right)\left( \partial_w \psi_{\bar{w}} + \partial_u \psi_{\bar{u}}\right) + f_-\left( \partial_w \psi_{\bar{u}} + \partial_u \psi_{\bar{w}} \right) + \frac{\bar{w}}{2v} \hat{\chi} = 0 \;, \label{2eome1e21}\\
& &\bar{\partial}_{\bar{w}} \chi + \frac{w}{v} \psi_{\bar{w}} = 0 \;, \label{2eome1e22}\\
& &\bar{\partial}_{\bar{u}} \chi + \frac{w}{v} \psi_{\bar{u}} = 0 \;. \label{2eome1e23}
\eea
The perturbative expansion is based on a zeroth order solution which solves the equations of motion for $f_+=f_-=0$ and reads \cite{Font:2009gq}
\bea
\chi^{(0)} &=& e^{-\frac{|w|^2}{\sqrt{2}v}} \;, \;\; \psi^{(0)}_{\bar{w}} = \frac{1}{\sqrt{2}} \chi^{(0)} \;,\;\;  \psi^{(0)}_{\bar{u}} = 0 \;.
\eea
As we regard this as a Higgs curve,
we are only concerned with a single solution so that the holomorphic function is just taken to be a constant.

The equations of motion are solved to ${\cal O}\left(z\right)$ by the ${\cal O}\left(z^2\right)$ expression
\bea
\chi &=& \chi^{(0)} \left[1 - \frac{1}{3\sqrt{2}v} \left( |w|^2 \left( \frac32 n_+u + \frac32 \bar{n}_+ \bar{u} + n_- w + \bar{n}_- \bar{w} \right) + \sqrt{2}v n_- w \right) + \chi^{(2)} \right] \;, \nn \\
\chi^{(2)} &=& w\left( \varphi_1 + u\varphi_2 + \bar{u}\varphi_3 + |u|^2\varphi_4 + u^2 \varphi_5 + \bar{u}^2 \varphi_6 \right) \;, \nn \\
\varphi_1 &=& \frac{n_-^2}{36v^2} w|w|^4 + \frac{|n_-|^2}{18v^2} \bar{w}|w|^4 + \frac{\bar{n}_-^2}{36v^2} \bar{w}^3|w|^2 + \frac{\sqrt{2}}{8v}\bar{w}|w|^2\left( \frac{5|n_+|^2}{4} - |n_-|^2\right) \nn \\
& & + \left( -\frac{\sqrt{2}n_-^2}{24v} +\frac{\sqrt{2}n_+^2}{16v}\right)w|w|^2 - \frac{\sqrt{2}}{8v}\bar{w}^3\left(\frac{7\bar{n}_-^2}{9} - \frac{\bar{n}_+^2}{2} \right) + \bar{w}\left( -\frac{5}{16}|n_+|^2 - \frac{|n_-|^2}{4} \right) \nn \\
& & - \left( \frac{n_-^2}{12} + \frac{n_+^2}{8} \right)w \;, \nn \\
\varphi_2 &=& \frac{1}{12v^2}|w|^4 n_+ n_- + \frac{1}{12v^2}\bar{w}^2|w|^2 n_+ \bar{n}_- - \frac{\sqrt{2}}{12v}|w|^2 n_+ n_- - \frac{\sqrt{2}}{4v}\bar{w}^2 \left( n_+ \bar{n}_- - \frac{1}{3}n_- \bar{n}_+ \right) - \frac{2}{3} n_+ n_- \;, \nn \\
\varphi_3 &=& \frac{1}{12v^2}|w|^4 \bar{n}_+ n_- + \frac{1}{12v^2}\bar{w}^2|w|^2 \bar{n}_+ \bar{n}_- - \frac{\sqrt{2}}{12v}|w|^2 \left(2\bar{n}_+ n_- - n_+ \bar{n}_- \right)- \frac{\sqrt{2}}{6v}\bar{w}^2 \bar{n}_+ \bar{n}_- \nn \\
& & - \frac{1}{3} \left(n_+ \bar{n}_- + n_-\bar{n}_+ \right)\;, \nn \\
\varphi_4 &=& \frac{|n_+|^2}{8v^2}\bar{w}|w|^2 - \frac{3\sqrt{2}}{8v}\bar{w}|n_+|^2 \;, \nn \\
\varphi_5 &=& \frac{n_+^2}{16v^2}\bar{w}|w|^2 - \frac{3\sqrt{2}n_+^2}{16v} \bar{w} \;, \nn \\
\varphi_6 &=& \frac{\bar{n}_+^2}{16v^2}\bar{w}|w|^2 - \frac{3\sqrt{2}\bar{n}_+^2}{16v} \bar{w} \;.
\eea
Like the matter curve solutions this solution involves $\bar{z}_1$ and $\bar{z}_2$ up to second order.

%%%%%%%%%%%%%%%%%%%%%%%%%%%%%%%%%%%%%%%%%%%%%%%%%%%%%%%%%%%%%%%%%%%%%%%%%%%%%
\subsection{Yukawa couplings}
\label{sec:yukcoupfinal}
%%%%%%%%%%%%%%%%%%%%%%%%%%%%%%%%%%%%%%%%%%%%%%%%%%%%%%%%%%%%%%%%%%%%%%%%%%%%%

Having calculated the perturbative wavefunctions it is possible to calculate the Yukawa couplings.
The computation of the physical Yukawa couplings requires both
the holomorphic Yukawa couplings and the normalisation of the kinetic terms.

The kinetic terms originate from terms in the 8d Lagrangian
$$
\int d^8 x -2 \omega \wedge D_{\mu} \bar{\psi} \wedge \bar{\sigma}^\mu \psi,
\qquad \int d^8 x D_{\mu} \chi \wedge \sigma^{\mu} \bar{\chi}.
$$
The contributions of these two terms should be combined to work out the normalisation for the actual zero mode.
Evaluation of the kinetic terms is not possible in a purely local framework, as it requires knowledge of the global
structure of the zero modes and in particular the full (compact) matter curve along which the zero mode is supported.
However it is important to note that the corrected wavefunctions induce inter-generational kinetic mixing since the corrections come in with powers of both $\bar{z}_{\bar{1}}$ and $\bar{z}_{\bar{2}}$.\footnote{This is in contrast to the flux corrected wavefunctions of \cite{Font:2009gq}.} Therefore to extract the physical Yukawa couplings requires a diagonalisation of the kinetic terms. The rank of the Yukawa matrix is unaltered by the kinetic terms and so we can analyse this without reference to the kinetic
terms.

The holomorphic Yukawas that appear in the superpotential are given by (\ref{yukcompute}). In \cite{Heckman:2008qa} a Frogatt-Nielsen-like mechanism was proposed for generating the observed mass hierarchies. The observation is that the integration volume form respects two symmetries $U(1)_1 \times U(1)_2$ given by
\be
z_1 \rightarrow e^{i\theta_1} z_1 \;, \\
z_2 \rightarrow e^{i\theta_2} z_2 \;.
\ee
The exponential part of the matter wavefunctions respects these symmetries and the exponential part of the Higgs wavefunctions breaks the symmetry down to the diagonal $U(1)$ given by
\be
\left(z_1 + z_2\right) \rightarrow e^{i\theta}\left(z_1 + z_2\right) \;.
\ee
It is this $U(1)$ that acts as a Frogatt-Nielsen-like charge since any integrals that do not respect it will vanish. This means that the non-exponential parts of the zero mode wavefunctions should combine in a $U(1)$ neutral way. For the top Yukawa coupling this occurs already at tree level. However since the other generations have non-trivial holomorphic functions $g\left(z_2,z_1\right)$ factors
 they can only form $U(1)$ neutral combinations through higher order non-holomorphic corrections to the wavefunctions.
 The idea is that since these are suppressed, they naturally induce a hierarchy.

 If such a hierarchy is present, we should be able to find it using the explicit solutions for the wavefunctions presented in sections
 \ref{sec:mattwav} and \ref{sec:higgswav}.
 Note that since we have only expanded the wavefunctions to second order in the derivative expansion we can only calculate the leading contributions to the Yukawa couplings $Y_{11}$, $Y_{12}$, $Y_{21}$, $Y_{22}$, $Y_{31}$ and $Y_{13}$.

The natural variables for calculating the Yukawas are $w$ and $u$ as introduced in eq. (\ref{lalala}).
The non vanishing integrals all take the form
\bea
I_{mn} &=& \int d^2z_1 d^2z_2 \;e^{-\frac{1}{v}\left(|z_1|^2+|z_2|^2+\frac{1}{\sqrt{2}}|z_1+z_2|^2\right)} |w|^{2m}|u|^{2n} \nn \\
&=&\int \frac14 d^2w d^2u \;e^{-\frac{1}{v}\left(\frac12 |u|^2+\frac{1}{2s}|w|^2\right)} |w|^{2m}|u|^{2n} \nn \\
&=&\pi^2 n!m!(2v)^{(n+m+2)}s^{(m+1)} \;,
\eea
where we define $s=\sqrt{2}-1$. Then calculating the integrals (\ref{yukcompute}) simply amounts to extracting the appropriate coefficient multiplying each $I_{mn}$ and summing the result. Such a calculation is naturally done using Mathematica.
Performing this calculation we find the following result
\bea
Y_{11} &=& -4a\pi^2v^2 \;, \\
Y_{ij} &=& 0 \;\; \forall\;(i,j)\neq(1,1)\;. \label{yukrank1}
\eea
We therefore find that flux and metric corrected wavefunctions do not induce new Yukawa couplings and
the Yukawa matrix remains as rank one.

This is a rather striking result given the form of the perturbed wavefunctions. It hints at a symmetry underlying the vanishing of the Yukawas. In a recent paper \cite{Cecotti:2009zf} precisely such a symmetry principle was shown. The idea is that the gauge symmetry of the 8D super-Yang-Mills can be used to define a kind of generalised cohomology where wavefunctions that differ by a gauge transformation are identified under the cohomology. A representative of each cohomology class was shown to be simply the wavefunctions evaluated on the localisation curve. These are just the holomorphic $g_i$ of (\ref{zewa}). Finally the Yukawa couplings were shown to depend just on the cohomology representatives and so could be evaluated simply by using the $g_i$. This directly gives our result (\ref{yukrank1}). Therefore (\ref{yukrank1}) is a special case of a more general phenomenon for all wavefunctions that are induced by corrections due to gauge flux.

%%%%%%%%%%%%%%%%%%%%%%%%%%%%%%%%%%%%%%%%%%%%%%%%%%%%%%%%%%%%%%%%%%%%%%%%%%%%%
\section{Summary and discussion}
\label{sec:summary}
%%%%%%%%%%%%%%%%%%%%%%%%%%%%%%%%%%%%%%%%%%%%%%%%%%%%%%%%%%%%%%%%%%%%%%%%%%%%%

In this article we have studied the inter-relationship of flavour and supersymmetry in the context of F-theory GUTs.
In the earlier part of the article we described the relationship between the topologically twisted theory of
\cite{Beasley:2008dc} and the canonical formulation of 8d super Yang-Mills theory, deriving the zero mode equations
of motion from the latter. We also provided a dictionary between the two formalisms. The simple scaling properties
of super Yang-Mills make it clear that metric rescalings (which include scalings of $\alpha_{GUT}$) should not alter the rank
of the Yukawa matrix. This is also required by holomorphy consideration of the 4d effective supergravity theory.

For metric and flux backgrounds we also solved the equations of motion up to second order in the perturbations.
We found that the induced metric deformations are the leading corrections to the wavefunctions dominating over the flux terms.
Using the explicit form for the wavefunctions in the vicinity of the triple intersection point we computed the
unnormalised Yukawa couplings. We found that the wavefunction deformations did not affect the rank of the
Yukawa matrix, which remained rank one. The fundamental reason for this has recently been explained in
\cite{Cecotti:2009zf}. The explicit form of the wavefunctions is however important
for non-holomorphic properties of the effective theory.

\subsection*{Acknowledgments}

We thank James Gray, Andre Lukas, Anshuman Maharana, Fernando Quevedo and Martijn Wijnholt for helpful discussions.
JC is supported by a Royal Society University Research Fellowship. EP is supported in
part by the European ERC Advanced Grant 226371 MassTeV, by the CNRS PICS no. 3059 and 4172,
by the grants ANR-05-BLAN-0079-02, the PITN contract PITN-GA-2009-237920 and was supported for a substantial
part of the work by an STFC Postdoctoral Fellowship.


\begin{thebibliography}{99}

%\cite{Donagi:2008ca}
\bibitem{Donagi:2008ca}
  R.~Donagi and M.~Wijnholt,
  ``Model Building with F-Theory,''
  arXiv:0802.2969 [hep-th].
  %%CITATION = ARXIV:0802.2969;%%

%\cite{Beasley:2008dc}
\bibitem{Beasley:2008dc}
  C.~Beasley, J.~J.~Heckman and C.~Vafa,
  ``GUTs and Exceptional Branes in F-theory - I,''
  JHEP {\bf 0901} (2009) 058
  [arXiv:0802.3391 [hep-th]].
  %%CITATION = JHEPA,0901,058;%%

%\cite{Hayashi:2008ba}
\bibitem{Hayashi:2008ba}
  H.~Hayashi, R.~Tatar, Y.~Toda, T.~Watari and M.~Yamazaki,
  ``New Aspects of Heterotic--F Theory Duality,''
  Nucl.\ Phys.\  B {\bf 806} (2009) 224
  [arXiv:0805.1057 [hep-th]].
  %%CITATION = NUPHA,B806,224;%%

  \bibitem{08052943}
  L.~Aparicio, D.~G.~Cerdeno and L.~E.~Ibanez,
  ``Modulus-dominated SUSY-breaking soft terms in F-theory and their test at
  %LHC,''
  JHEP {\bf 0807} (2008) 099
  [arXiv:0805.2943 [hep-ph]].
  %%CITATION = JHEPA,0807,099;%%

%\cite{Beasley:2008kw}
\bibitem{Beasley:2008kw}
  C.~Beasley, J.~J.~Heckman and C.~Vafa,
  ``GUTs and Exceptional Branes in F-theory - II: Experimental Predictions,''
  JHEP {\bf 0901} (2009) 059
  [arXiv:0806.0102 [hep-th]].
  %%CITATION = JHEPA,0901,059;%%

%\cite{Donagi:2008kj}
\bibitem{Donagi:2008kj}
  R.~Donagi and M.~Wijnholt,
  ``Breaking GUT Groups in F-Theory,''
  arXiv:0808.2223 [hep-th].
  %%CITATION = ARXIV:0808.2223;%%

%\cite{Heckman:2008qt}
\bibitem{Heckman:2008qt}
  J.~J.~Heckman and C.~Vafa,
  ``F-theory, GUTs, and the Weak Scale,''
  arXiv:0809.1098 [hep-th].
  %%CITATION = ARXIV:0809.1098;%%

%\cite{Blumenhagen:2008zz}
\bibitem{Blumenhagen:2008zz}
  R.~Blumenhagen, V.~Braun, T.~W.~Grimm and T.~Weigand,
  ``GUTs in Type IIB Orientifold Compactifications,''
  Nucl.\ Phys.\  B {\bf 815} (2009) 1
  [arXiv:0811.2936 [hep-th]].
  %%CITATION = NUPHA,B815,1;%%

%\cite{Font:2008id}
\bibitem{Font:2008id}
  A.~Font and L.~E.~Ibanez,
  ``Yukawa Structure from U(1) Fluxes in F-theory Grand Unification,''
  JHEP {\bf 0902} (2009) 016
  [arXiv:0811.2157 [hep-th]].
  %%CITATION = JHEPA,0902,016;%%

%\cite{Heckman:2008qa}
\bibitem{Heckman:2008qa}
  J.~J.~Heckman and C.~Vafa,
  ``Flavor Hierarchy From F-theory,''
  arXiv:0811.2417 [hep-th].
  %%CITATION = ARXIV:0811.2417;%%

%\cite{Blumenhagen:2008aw}
\bibitem{Blumenhagen:2008aw}
  R.~Blumenhagen,
  ``Gauge Coupling Unification In F-Theory Grand Unified Theories,''
  Phys.\ Rev.\ Lett.\  {\bf 102} (2009) 071601
  [arXiv:0812.0248 [hep-th]].
  %%CITATION = PRLTA,102,071601;%%

\bibitem{09013785}
  J.~L.~Bourjaily,
  ``Local Models in F-Theory and M-Theory with Three Generations,''
  arXiv:0901.3785 [hep-th].
  %%CITATION = ARXIV:0901.3785;%%

%\cite{Hayashi:2009ge}
\bibitem{Hayashi:2009ge}
  H.~Hayashi, T.~Kawano, R.~Tatar and T.~Watari,
  ``Codimension-3 Singularities and Yukawa Couplings in F-theory,''
  Nucl.\ Phys.\  B {\bf 823} (2009) 47
  [arXiv:0901.4941 [hep-th]].
  %%CITATION = NUPHA,B823,47;%%

\bibitem{09024143}
  B.~Andreas and G.~Curio,
  ``From Local to Global in F-Theory Model Building,''
  arXiv:0902.4143 [hep-th].
  %%CITATION = ARXIV:0902.4143;%%

  \bibitem{09033009}
  C.~M.~Chen and Y.~C.~Chung,
  %``A Note on Local GUT Models in F-Theory,''
  Nucl.\ Phys.\  B {\bf 824} (2010) 273
  [arXiv:0903.3009 [hep-th]].
  %%CITATION = NUPHA,B824,273;%%

%\cite{Donagi:2009ra}
\bibitem{Donagi:2009ra}
  R.~Donagi and M.~Wijnholt,
  ``Higgs Bundles and UV Completion in F-Theory,''
  arXiv:0904.1218 [hep-th].
  %%CITATION = ARXIV:0904.1218;%%

%\cite{Bouchard:2009bu}
\bibitem{Bouchard:2009bu}
  V.~Bouchard, J.~J.~Heckman, J.~Seo and C.~Vafa,
  ``F-theory and Neutrinos: Kaluza-Klein Dilution of Flavor Hierarchy,''
  arXiv:0904.1419 [hep-ph].
  %%CITATION = ARXIV:0904.1419;%%

%\cite{Randall:2009dw}
\bibitem{Randall:2009dw}
  L.~Randall and D.~Simmons-Duffin,
  ``Quark and Lepton Flavor Physics from F-Theory,''
  arXiv:0904.1584 [hep-ph].
  %%CITATION = ARXIV:0904.1584;%%

  \bibitem{09043101}
  J.~J.~Heckman and C.~Vafa,
  ``CP Violation and F-theory GUTs,''
  arXiv:0904.3101 [hep-th].
  %%CITATION = ARXIV:0904.3101;%%

\bibitem{09043932}
  J.~Marsano, N.~Saulina and S.~Schafer-Nameki,
  ``F-theory Compactifications for Supersymmetric GUTs,''
  JHEP {\bf 0908} (2009) 030
  [arXiv:0904.3932 [hep-th]].
  %%CITATION = JHEPA,0908,030;%%

\bibitem{09052289}
  R.~Tatar, Y.~Tsuchiya and T.~Watari,
  ``Right-handed Neutrinos in F-theory Compactifications,''
  Nucl.\ Phys.\  B {\bf 823} (2009) 1
  [arXiv:0905.2289 [hep-th]].
  %%CITATION = NUPHA,B823,1;%%

\bibitem{09060013}
  R.~Blumenhagen, T.~W.~Grimm, B.~Jurke and T.~Weigand,
  ``F-theory uplifts and GUTs,''
  JHEP {\bf 0909} (2009) 053
  [arXiv:0906.0013 [hep-th]].
  %%CITATION = JHEPA,0909,053;%%

%\cite{Marsano:2009gv}
\bibitem{Marsano:2009gv}
J.~Marsano, N.~Saulina and S.~Schafer-Nameki,
``Monodromies, Fluxes, and Compact Three-Generation F-theory GUTs,''
arXiv:0906.4672 [hep-th].
%%CITATION = ARXIV:0906.4672;%%

%\cite{Heckman:2009mn}
\bibitem{Heckman:2009mn}
  J.~J.~Heckman, A.~Tavanfar and C.~Vafa,
  ``The Point of E8 in F-theory GUTs,''
  arXiv:0906.0581 [hep-th].
  %%CITATION = ARXIV:0906.0581;%%

\bibitem{09063297}
  R.~Blumenhagen, J.~P.~Conlon, S.~Krippendorf, S.~Moster and F.~Quevedo,
  ``SUSY Breaking in Local String/F-Theory Models,''
  JHEP {\bf 0909} (2009) 007
  [arXiv:0906.3297 [hep-th]].
  %%CITATION = JHEPA,0909,007;%%

%\cite{Conlon:2009qa}
\bibitem{Conlon:2009qa}
  J.~P.~Conlon and E.~Palti,
  ``On Gauge Threshold Corrections for Local IIB/F-theory GUTs,''
  arXiv:0907.1362 [hep-th].
  %%CITATION = ARXIV:0907.1362;%%

%\cite{Font:2009gq}
\bibitem{Font:2009gq}
  A.~Font and L.~E.~Ibanez,
  ``Matter wave functions and Yukawa couplings in F-theory Grand Unification,''
  arXiv:0907.4895 [hep-th].
  %%CITATION = ARXIV:0907.4895;%%

%\cite{Blumenhagen:2009yv}
\bibitem{Blumenhagen:2009yv}
  R.~Blumenhagen, T.~W.~Grimm, B.~Jurke and T.~Weigand,
  ``Global F-theory GUTs,''
  arXiv:0908.1784 [hep-th].
  %%CITATION = ARXIV:0908.1784;%%


\bibitem{08070789}
  J.~P.~Conlon, A.~Maharana and F.~Quevedo,
  ``Wave Functions and Yukawa Couplings in Local String Compactifications,''
  JHEP {\bf 0809} (2008) 104
  [arXiv:0807.0789 [hep-th]].
  %%CITATION = JHEPA,0809,104;%%


%\cite{Cecotti:2009zf}
\bibitem{Cecotti:2009zf}
  S.~Cecotti, M.~C.~N.~Cheng, J.~J.~Heckman and C.~Vafa,
  ``Yukawa Couplings in F-theory and Non-Commutative Geometry,''
  arXiv:0910.0477 [hep-th].
  %%CITATION = ARXIV:0910.0477;%%

\end{thebibliography}
\end{document}